\documentclass[aps,preprint,amsfonts,amsmath,amssymb,superscriptaddress]{revtex4-2}

\pdfoutput=1

\usepackage{hyperref}
\usepackage[utf8x]{inputenc}	
\usepackage[T1]{fontenc}
\usepackage{mathrsfs}
\usepackage{color}
\usepackage{fancyhdr}
\usepackage{verbatim}
\usepackage{graphicx}
\usepackage{amsmath,amssymb}
\usepackage[]{natbib}

\usepackage[title]{appendix}
\usepackage{placeins} 

\usepackage{caption}
\usepackage{subcaption}
\usepackage{floatrow}

\begin{document}

\title{Innovative Electroacoustic resonator Control enforcing Duffing dynamics at moderate excitation levels: conception and experimental validation.}
\author{Emanuele De Bono\footnote{emanuele.de-bono@ec-lyon.fr}}
\affiliation{LTDS, \'Ecole Centrale de Lyon -- CNRS UMR5513, 36 avenue Guy de Collongue, 69134 Ecully Cedex (France)}
\author{Maxime Morell}
\affiliation{Univ Lyon,  Ecole Nationale des Travaux Publics de l’Etat, LTDS UMR 5513, F-69518 Vaulx en Velin Cedex (France)}
\author{Manuel Collet}
\affiliation{LTDS, \'Ecole Centrale de Lyon -- CNRS UMR5513, 36 avenue Guy de Collongue, 69134 Ecully Cedex (France)}

\author{Emmanuel Gourdon}
\affiliation{Univ Lyon,  Ecole Nationale des Travaux Publics de l’Etat, LTDS UMR 5513, F-69518 Vaulx en Velin Cedex (France)}
\author{Alireza Ture Savadkoohi}
\affiliation{Univ Lyon,  Ecole Nationale des Travaux Publics de l’Etat, LTDS UMR 5513, F-69518 Vaulx en Velin Cedex (France)}

\author{Claude Henri Lamarque}
\affiliation{Univ Lyon,  Ecole Nationale des Travaux Publics de l’Etat, LTDS UMR 5513, F-69518 Vaulx en Velin Cedex (France)}


\begin{abstract}
	The electroacoustic resonator is an efficient electro-active device for noise attenuation in enclosed cavities or acoustic waveguides. It is made of a loudspeaker (the actuator) and one or more microphones (the sensors). So far, the desired acoustic behaviour, expressed in terms of a linear relationship between sound pressure and vibrational motion, has been more efficiently achieved by a model-inversion strategy which is implemented by driving the electrical current in the loudspeaker coil, based upon the measured pressure. The corrector transfer function is hence digitally executed by the classical infinite-impulse-response technique. In order to enforce non-linear, instead of linear, operators between the pressure and vibrational motion of the speaker diaphragm, in the same pressure-based, current-driven architecture, the transfer-function-based control strategies must be abandoned. In this manuscript, we present a novel technique based upon real-time integration of potentially any target dynamics (either linear or non-linear). Then, we experimentally validate such control strategy by enforcing a Duffing type dynamics at excitation levels far below the linearity limit of the mechano-acoustical dynamics of the resonator. Such control strategy demonstrates to efficiently reproduce nonlinear operators, opening the doors for the experimental investigation of noise-control by a potentially vast spectrum of programmable non-linear boundaries, under the ordinary sound excitation levels.
\end{abstract}

\maketitle

\section{Introduction}\label{sec:introduction}

The noise-control by treating the boundaries of propagative domains is a large area of research encompassing all fields from electromagnetics to solid mechanics and acoustics. In acoustics, a typical boundary treatment problem is the room modal equalization, where the objective is to damp the acoustic modes in an enclosed cavity. P. Morse \cite{morse1939some}, in 1939, recognized the normal surface impedance as the quantity characterizing the acoustic behaviour of a locally reacting boundary, in its \emph{linear regime}. It is defined as the ratio between the local sound pressure and the normal velocity in the complex space: $Z_s(s)=p(s)/v(s)$, where $s$ is the Laplace variable, equal to $\mathrm{j}\omega$ in the stationary regime. However, a generic boundary might present non-locally reacting and/or non-linear acoustic behaviour, and can be characterized by a general operator $\mathcal{L}\{p(\mathbf{x},t),\mathbf{u}(\mathbf{x},t)\}=0$ relating sound pressure $p$ and the surface displacement $\mathbf{u}$ in the physical coordinates $(\mathbf{x}$, $t)$. In case of locally-reacting and linear behaviour of the boundary, the implicit general operator $\mathcal{L}\{\bullet\}$ degenerates to a linear relationship between local sound pressure $p(t)$ and normal velocity $v(t)$, whose Laplace or Fourier transform leads to the definition of the normal surface impedance: $p(s)-Z_s(s)v(s)=0$.\\
Though the linear regime is valuable only below a certain threshold of the involved energy, both in acoustics and solid mechanics the problem of noise and vibration mitigation has been mostly tackled by \emph{linear} means. In solid mechanics, the Tuned-Vibration-Absorber (TVA), consisting of a resonator attached to the main structure, properly tuned with the resonance of the primary structure (supposed linear), has more than a century tradition \cite{Frahm1909} and established industrial practice. Nevertheless, since mistuned TVA might increase the vibration level of the primary structure, \emph{adaptive} TVA absorbers have been developed with controllable or adjustable parameters, as well as \emph{active} TVAs \cite{sun1995passive}. The latter, thanks to an \emph{active} force on the mass, can provide broader bandwidth, and when the active element fails, they can still function as passive vibration absorbers (fail-safe).\\
The resonance principle of energy capture is also exploited in sound mitigation by the Helmholtz or quarter-wavelength resonators, which are the equivalent in fluid acoustics of the TVA in solid mechanics. Their main drawback is related to the fact that their equivalent acoustical stiffness term is related to the air compressibility in their acoustic cavity. Hence, to target lower frequencies, larger volumes of air are required, limiting their implementation when confronted with space and weight strict specifications as in aeronautics for example \cite{ma2020development}. Also the Helmholtz and quarter-wavelength resonators have found some evolution by \emph{adaptive} solutions for coping with tunability requirements \cite{ma2020development}. Other classical passive acoustic resonators are membranes, which are also exploited in (electro)-\emph{active} devices, such as the so-called Electro-Acoustic absorbers (EA). As in the active TVA, a \emph{control} force is applied to the mass of the EA which can modify its response to the acoustic excitation. Most commonly used technology for the EA is the loudspeaker, where the speaker membrane dynamics is altered by the Lorentz/Laplace control force. From the seminal idea of Olson and May \cite{olson1953electronic}, the EA concept has given rise to various strategies, such as electrical-shunting \cite{fleming2007control}, direct-impedance control \cite{furstoss1997surface} and self-sensing \cite{leo2000self}. In order to overcome the low-flexibility drawback of electrical shunting techniques, minimize the number of sensors, meanwhile avoiding to get involved into the electrical-inductance modelling of the loudspeaker, a pressure-based current-driven architecture proved to achieve the best absorption performances in terms of both bandwidth and tunability \cite{rivet2016broadband}. It employs one or more pressure sensors (microphones) nearby the speaker, and a model-inversion technique to target the desired impedance by controlling the electrical current in the speaker coil. 
Its efficiency has been demonstrated for room-modal equalization \cite{rivet2016room}, sound transmission mitigation in waveguides \cite{boulandet2018duct}, \cite{billon2021experimental}, and even non-reciprocal propagation \cite{KarkarDeBono2019}.\\
In solid mechanics, it has long been tried to overcome the limits of linear TVA by exploiting non-linear (typically Duffing type) resonators \cite{roberson1952synthesis}, \cite{jordanov1988optimal}. More recently, it has been demonstrated that the coupled dynamics of the linear main structure and non-linear absorber, can feature a special phenomenon: the Target Energy Transfer (TET) where vibrational energy is transferred from the linear host structure to the non-linear absorber in a one-way and irreversible fashion \cite{aubry2001analytic}, \cite{vakakis2008nonlinear}. We speak in this case of a Nonlinear-Energy-Sink (NES), which presents great interest for vibration quick suppression and energy harvesting, and whose potentialities are still under investigation. Widely studied in solid mechanics, the nonlinear absorption potentials are recently getting explored also in acoustics. An extensive work in this direction has been realized by the team of the Laboratoire de Mécanique et d'Acoustique de Marseille, in France, since \cite{cochelin2006experimental}, where they achieved TET from a linear acoustic cavity to a \emph{weakly-coupled} thin visco-elastic membrane which behaves as a Duffing resonator. As the linear ones, passive nonlinear absorbers are not easily tunable for targeting different bandwidths. Moreover, they usually need high-energy threshold in order to trigger the nonlinear behaviour. An electro-active nonlinear absorber might overcome these limitations, by \emph{transforming} the mechano-acoustical dynamics of the loudspeaker from linear to nonlinear, while keeping the same external excitation levels. In order to do that, in \cite{guo2020improving}, an additional microphone was placed inside the EA back-cavity, such that to retrieve a measurement directly proportional to the diaphragm displacement at low frequencies. The nonlinear behaviour was then induced by defining an electrical current (the controller) which comprises a ``linear'' ($i_L$) and a ``nonlinear'' ($i_{NL}$) contribution separately. The first one ($i_L$) is in charge of inverting the loudspeaker own dynamics and enforcing the linear part of the target dynamics, while the second one ($i_{NL}$) is in charge of enforcing the nonlinear term, proportional to the cubic pressure inside the back-cavity of the EA. To apply this strategy, attention must be put in the recursive definition of the ``linear'' term $i_L$ in its digital implementation. In the classical Infinite-Impulse-Response (IIR) recursive scheme, the controller at each time step is defined based upon the controller at the previous time steps. Hence, in order for $i_L$ to correctly accomplish the model-inversion and the ``linear'' dynamics targeting, in the IIR scheme, $i_L$ should never be mixed up with $i_{NL}$ in \cite{guo2020improving}. If the model inversion is achieved by IIR, it is never possible to target a purely (so-called \emph{essential} \cite{nayfeh1980nonlinear}) nonlinear behaviour of the EA. Indeed, in \cite{guo2020improving}, that which is defined as ``nonlinear impedance control'' actually combines the passive EA behaviour (which is linear) with a cubic essential nonlinearity. 
The nonlinear behaviour accomplished by this device, manifesting itself in the \emph{hardening} or \emph{softening} spring effects, is mildly evident as a small enlargement of the bandwidth with respect to the purely linear control. The great achievement of \cite{guo2020improving} though, was to be able of enforcing a nonlinear acousto-mechanical dynamics of the EA at sufficiently moderate excitation levels. This was accomplished by increasing the gain multiplying the sensed pressure in the back-cavity, which was however limited by important stability constraints.\\
In this paper we propose an innovative strategy to enforce nonlinear mechano-acoustical dynamics of the EA, in the same pressure-based, current-driven architecture of \cite{rivet2016broadband}, without additional sensors to estimate the motion of the speaker membrane. The idea is to keep using the model-inversion technique to target nonlinear, instead of linear, operators. As the target dynamics is non-linear, the control function cannot be written as a convolution system, and the classical digital implementation techniques (such as the IIR) based upon approximation of the discrete convolution, had to be abandoned in favour of \emph{real-time integration} schemes.\\
In Section \ref{sec:conception}, the control algorithm is outlined, while in Section \ref{sec:experimental}, experimental tests validate the achievement of a tunable Duffing dynamics at moderate excitation levels. Finally, in Section \ref{sec:conclusions}, we give the conclusions and future developments.

\section{Conception}\label{sec:conception}

This control strategy makes use of the same model-inversion as in \cite{rivet2016broadband}, but without passing by the Laplace transform. In order to enforce a possibly nonlinear term in a variable (the membrane motion) which is not measured (either directly or indirectly, as instead done in \cite{guo2020improving}), the model inversion is here employed directly at each time step, so that the actual dynamics of the actuator (the speaker) would match a \emph{target} displacement evolution. This latter is obtained by integrating in real-time the desired dynamics equation relating the \emph{targeting} displacement and the measured pressure. We first delineate the in-time model-inversion idea in Section \ref{sec:model inversion}, and then the complete Real-Time-Integration (RTI) control algorithm is exposed in Section \ref{sec:algorithm}.

\subsection{The SDOF model inversion}\label{sec:model inversion}

\begin{figure}
	\centering
	\includegraphics[width=.6\textwidth]{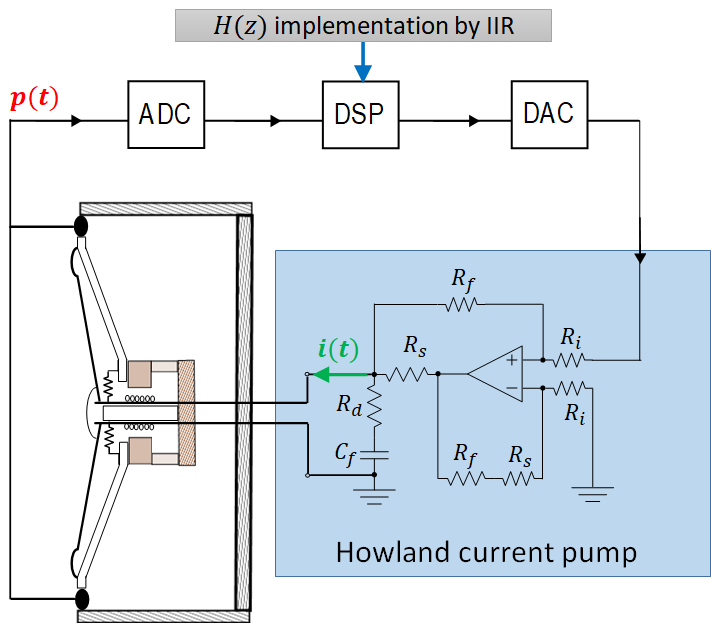}
	\caption{Sketch of the EA architecture.}
	\label{fig:EA_with_Howland}
\end{figure}

The model inversion idea is to fix a desired output behaviour, and to define the input based upon the knowledge of the \emph{system model} \cite{devasia1996nonlinear}. For a linear, SDOF system model, the model inversion is banal.\\
In \cite{rivet2016broadband}, model inversion is employed to target a certain desired impedance on the diaphragm of a loudspeaker, by means of a pressure-based, current-driven control architecture. The system is described in Figure \ref{fig:EA_with_Howland}: the pressure $p$ on the speaker diaphragm estimated by one or more microphones around the membrane, after being digitally converted by the Analogue-Digital-Converter (ADC), is fed into a \emph{programmable} digital signal processor (DSP) where the corrector transfer function $H(z)$ in the discrete Laplace variable $z$, is implemented by the IIR scheme. The Howland current pump \cite{pease2008comprehensive} allows to enforce the electrical current $i$ in the speaker coil independently of the voltage at the loudspeaker terminals. It consists of an operational amplifier, two input resistors $R_i$, two feedback resistors $R_f$, and a current sense resistor $R_s$. The resistance $R_d$ and capacitance $C_f$ constitutes the compensation circuit to ensure stability with the grounded load \cite{steele1992tame}.

\begin{figure}
	\centering
	\includegraphics[width=.9\textwidth]{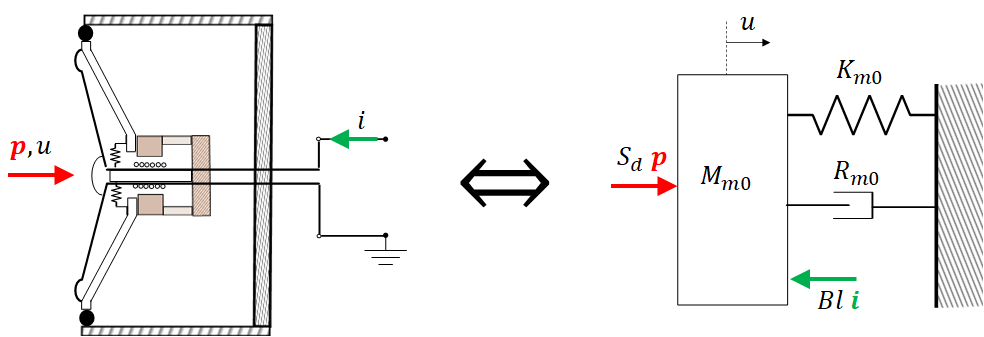}
	\caption{Sketch of the SDOF piston-mode system employed for the model inversion.}
	\label{fig:EA_model}
\end{figure}

The corrector transfer function needed to achieve a certain target acoustic impedance $Z_{at}(s)$, is synthesized based upon the SDOF piston-mode approximation of the loudspeaker mechano-acoustical dynamics. This latter is reported in Eq. \ref{eq:LS mech dynamics in time} in the time domain in terms of the normal displacement $u(t)$ of the diaphragm, and is illustrated in Figure \ref{fig:EA_model}.

\begin{equation}
	\label{eq:LS mech dynamics in time}
	M_{a0} \ddot{u}(t) + R_{a0} \dot{u}(t) + K_{a0} u(t) = p(t) - \frac{Bl}{S_d}\; i(t), \\
\end{equation}

In Eq. \ref{eq:LS mech dynamics in time}, $M_{a0}$, $R_{a0}$ and $K_{a0}$ are the acoustical mass, resistance and stiffness of the SDOF loudspeaker model in case of zero electrical current, $S_d$ is the so-called equivalent piston area corresponding to the participation factor of the pressure field on the piston mode, and $Bl$ is the so-called force-factor, corresponding to the magnetic field times the length of the coil.\\
In the Laplace domain, Eq. \ref{eq:LS mech dynamics in time} writes:

\begin{equation}
	\label{eq:LS mech dynamics Laplace}
	Z_{a0}(s) v(s)  = p(s) - \frac{Bl}{S_d}\; i(s), \\
\end{equation}

where $v(s)$ is the normal velocity and $Z_{a0}(s) = M_{a0} s + R_{a0} + \frac{K_{a0}}{s}$ the acoustical impedance of the loudspeaker in open circuit (with $i(s)=0$).\\

Based upon the model of Eq. \eqref{eq:LS mech dynamics Laplace}, it is easy to derive the transfer function $H(s)=i(s)/p(s)$ in order to achieve the desired impedance $p(s)/v(s)=Z_{at}(s)$:

\begin{equation}
	\label{eq:H(s)}
	H(s) = \frac{i(s)}{p(s)}= \frac{S_d}{Bl} \biggr(1 - \frac{Z_{a0}(s)}{Z_{at}(s)}\biggr) 
\end{equation} 

For its digital implementation, the corrector transfer function $H(s)$ is transformed in the discrete Laplace space $z$ by a zero-order-holder (zoh) or Tustin transform. Such transformations allow to approximate the convolution integral $i(t)=\int_{0}^{t}H(t)p(t-\tau)d\tau$ by a finite sum of lower rectangles (zoh) or trapezoids (Tustin), hence to obtain the controller $i(t)$ at each time step.
This is the model-inversion control designed by \cite{rivet2016broadband}.\\
In view of the real-time integration strategy outlined in the next section, it is useful to rewrite the controller $i(s)$ in terms of the target velocity $v_{at}(s)=p(s)/Z_{at}(s)$:

\begin{equation}
	\label{eq:H(s) with v_at}
	i(s) = H(s)p(s) = \frac{S_d}{Bl} \biggr( p(s) - Z_{a0}(s)\; v_{at}(s)\biggr).
\end{equation}

The exact inverse Laplace transform of Eq. \eqref{eq:H(s) with v_at}, gives:

\begin{equation}
	\label{eq:i(t) model inversion}
	\begin{split}
		i(t)=&\frac{S_d}{Bl}\biggr[ p(t)- \biggr(M_{a0}\ddot{u}_{at}(t) + R_{a0}\dot{u}_{at}(t) + K_{a0}u_{at}(t)\biggr) \biggr],
	\end{split}
\end{equation}

where $u_{at}(t)$ is the target displacement in time domain. Supposing to have assigned the time history of $u_{at}(t)$, implementing the controller of Eq. \eqref{eq:i(t) model inversion} in the dynamics model of Eq. \eqref{eq:LS mech dynamics in time}, leads to the following equation:

\begin{equation}
	\label{eq:equality}
	\begin{split}
		M_{a0}\biggr(\ddot{u}(t) - \ddot{u}_{at}(t)\biggr) + R_{a0}\biggr(\dot{u}(t) - \dot{u}_{at}(t)\biggr) + K_{a0}\biggr(u(t) - u_{at}(t)\biggr) = 0.
	\end{split}
\end{equation}

Eq. \eqref{eq:equality} is a homogenous second-order linear differential equation in the variable $d(t)=u(t)-u_{at}(t)$ with positive coefficients. If $d(0)\neq0$ or $\dot{d}(0)\neq0$, the solution $d(t)$ exponentially decays to zero, hence after a short transient the actual dynamics follows the target one, i.e. $u(t)=u_{at}(t)$. Observe that Eq. \eqref{eq:i(t) model inversion} is the model-inverting controller in time domain, written in terms of the target displacement $u_{at}(t)$, rather than in terms of the target impedance. Hence, it can be employed for implementing a target displacement relative to a nonlinear target operator. In the next section, we describe the control strategy to enforce a general (possibly nonlinear) mechano-acoustical operator, exploiting the model-inversion concept in time domain which, for the SDOF piston-mode approximation of a loudspeaker, writes as in Eq. \eqref{eq:i(t) model inversion}.

\subsection{The RTI algorithm}\label{sec:algorithm}

In this section, the RTI algorithm is described considering a general target differential operator and an equally general actuator dynamics, both written in the system state-space.\\
So let us suppose to target a general local dynamics, corresponding to a possibly nonlinear differential equation or order $N$. It can be written in its state-space representation as: 

\begin{equation}
	\label{eq:state-space operator}
	\begin{split}
		\mathbf{\dot{y}_{at}}(t)=\mathbf{f}(\mathbf{y_{at}}(t),p(t)),
	\end{split}
\end{equation}

where $\mathbf{y}_{at}$ is the \emph{target} state vector containing the target displacement $u_{at}(t)$ and its time-derivatives up to the order ($N-1$), and $\mathbf{\dot{y}_{at}}$ is its time derivative. Observe that the acoustically excited EA is, in general, a nonideal system \cite{nayfeh1980nonlinear}, in the sense that the acoustic pressure $p$ depends upon the response of the system itself $\mathbf{y_{at}}$. In a non-perfectly anechoic environment indeed, the incident and scattered pressure fields are coupled between each others. The coupling operator depends upon the external acoustic environment (so-called acoustic load in 1D). Since the EA objective is to operate without prior knowledge of the cavity wherein it is placed (see Section \ref{sec:introduction}), such coupling between incident and scattered fields is not known a-priori, therefore the need for measuring the sound pressure. For this reason, on the one hand $p$ is considered as a separate variable with respect to $\mathbf{y_{at}}$ in Eq. \eqref{eq:state-space operator} (though actually physically coupled in enclosed cavities) and, on the other hand, it is natural to include $p$ in the function $\mathbf{f}$, and do not consider it as an external excitation term.\\
Let us also consider a general dynamics of the actuator, expressed in terms of the state space variables:

\begin{equation}
	\label{eq:actuator dynamics}
	\begin{split}
		g\biggr( \mathbf{\dot{y}}(t), \mathbf{y}(t), p(t) \biggr) - i(t) = 0,
	\end{split}
\end{equation}

where $i(t)$ is the control variable, $\mathbf{y}(t)$ is the \emph{actual} state vector (containing the actual displacement $u(t)$ and its time-derivatives up to the order $N-1$) and  $g\biggr( \mathbf{\dot{y}}(t), \mathbf{y}(t), p(t) \biggr)$ is the function describing the dynamics of the actuator in its passive mode (with $i(t)=0$).\\
As said above, the algorithm used to enforce the target dynamics described by Eq. \eqref{eq:state-space operator} in real time, essentially consists of two operations, to be carried out at each time step $t_n$:\\

Time step $t_n$:

\begin{enumerate} 
	
	\item Evaluate $\mathbf{\dot{y}_{at}}(t_n)=\mathbf{f}(\mathbf{y_{at}}(t_n),p(t_n))$, from the measured pressure $p(t_n)$ and the state vector $\mathbf{y_{at}}(t_n)$ estimated at the previous time step. Then, inject the controller:
	
	\begin{equation}
		\label{eq:model-inverting controller general actuator}
		\begin{split}
			i(t_n) = g_{est}\biggr( \mathbf{\dot{y}_{at}}(t_n), \mathbf{y_{at}}(t_n), p(t_n) \biggr),
		\end{split}
	\end{equation}
	
	where $g_{est}(\bullet)$ is the model estimating the actual dynamics $g(\bullet)$ of the actuator.
	
	\item Apply an integration scheme, such as Runge-Kutta of $4^{th}$ order (RK4), to   $\mathbf{f}(\mathbf{y_{at}(t_n)},p(t_n))$ and estimate the target state vector $\mathbf{y_{at}(t_{n+1})}$ for the next step. 
	
	%
	%
	%
	
\end{enumerate}

Then restart from point 1. with $t_n=t_{n+1}$.\\

Injecting the controller defined by Eq. \eqref{eq:model-inverting controller general actuator} in the actuator dynamics Eq. \eqref{eq:actuator dynamics} at each time step $t=t_n$, produces:

\begin{equation}
	\label{eq:equality general actuator}
	\begin{split}
		g\biggr( \mathbf{\dot{y}}(t_n), \mathbf{y}(t_n), p(t_n) \biggr) = g_{est}\biggr( \mathbf{\dot{y}_{at}}(t_n), \mathbf{y_{at}}(t_n), p(t_n) \biggr).
	\end{split}
\end{equation}

As long as $g_{est}(\bullet)$ is a sufficiently accurate estimation of $g(\bullet)$, Eq. \eqref{eq:equality general actuator} yields $\mathbf{y}(t_n)=\mathbf{\dot{y}_{at}}(t_n)$ after a short transient. For the loudspeaker actuator, Eq. \eqref{eq:equality general actuator} becomes the Eq. \eqref{eq:equality}.\\

Such algorithm poses no restriction, a priori, on the target dynamics, and its accuracy mainly depends upon the size of the sampling period which affects the precision and stability of the integration scheme, as well as on accuracy of the system model identification.\\ 
In the next section, the RTI control strategy is implemented experimentally to enforce both a linear and a Duffing mechano-acoustical dynamics on an EA prototype. 

\section{Experimental validation}\label{sec:experimental}

\begin{figure}[ht!]
	\centering
	\includegraphics[width=0.8\textwidth]{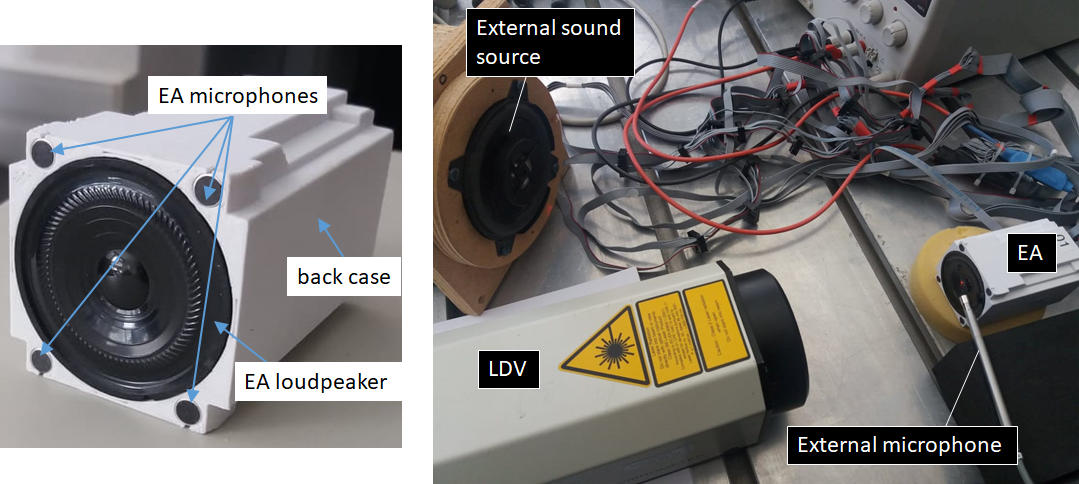}
	\caption{Photos of the EA prototype (left) and of the experimental test-rig (right) for the measurement of pressure and velocity on the speaker diaphragm..}
	\label{fig:Experimental photos}
\end{figure}

In this chapter, the RTI algorithm is experimentally implemented to target a linear (in Section \ref{sec:linear target dynamics}) and a nonlinear dynamics (in Section \ref{sec:nonlinear target dynamics}) of the EA. The system model to be ``inverted'' is taken as the SDOF piston-mode approximation of the speaker mechano-acoustical dynamics, as in Section \ref{sec:model inversion}. Both the signal acquisition and the control implementation is operated by a D-Space MicroLaBox DS1202 hardware. The EA prototype employed for the experimental validation is photographed in Figure \ref{fig:Experimental photos} on the left. It consists of a central speaker with four corner microphones used to retrieve the averaged pressure on the speaker. The back-case accommodates the analogical electronic card interfacing with the D-Space. The test-bench to measure pressure and velocity on the speaker diaphragm is illustrated in Figure \ref{fig:Experimental photos} on the right. An external acoustic source excites the EA, whose dynamics is retrieved in terms of sound pressure (by an external Brüel and Kjaer microphone) and normal velocity at the centre of the speaker diaphragm (by a Laser-Doppler-Velocimeter LDV).\\
The parameters of the loudspeaker SDOF model employed in our controller (Eq. \eqref{eq:i(t) model inversion}), are reported in Table \ref{tab:TSparam FEMTO}. They have been estimated by impedance measurements in different configurations, as reported in \cite{DeBono2021a}.

\begin{table}
	\centering
	\begin{tabular}{c|cccc}
		\hline
		\hline
		Thiele-Small parameters & $M_{a0}$ & $R_{a0}$ & $K_{a0}$ & $Bl/S_d$\\
		\hline
		Units & $\mathrm{kg.m^{-2}}$ & $\mathrm{Pa.s.m^{-1}}$ & $\mathrm{Pa.m^{-1}}$ & $\mathrm{Pa.A^{-1}}$ \\
		\hline
		Values & $4.45\times10^{-4}$ & $0.173$ & $3.85\times10^{3}$ & $1.10$ \\
		\hline
		\hline
	\end{tabular}
	\caption{Thiele-Small parameters of the EA.}
	\label{tab:TSparam FEMTO}
\end{table}

\subsection{Linear target dynamics}\label{sec:linear target dynamics}

\begin{figure}[ht!]
	\centering
	\begin{subfigure}[ht!]{0.45\textwidth}
		\centering
		\includegraphics[width=\textwidth]{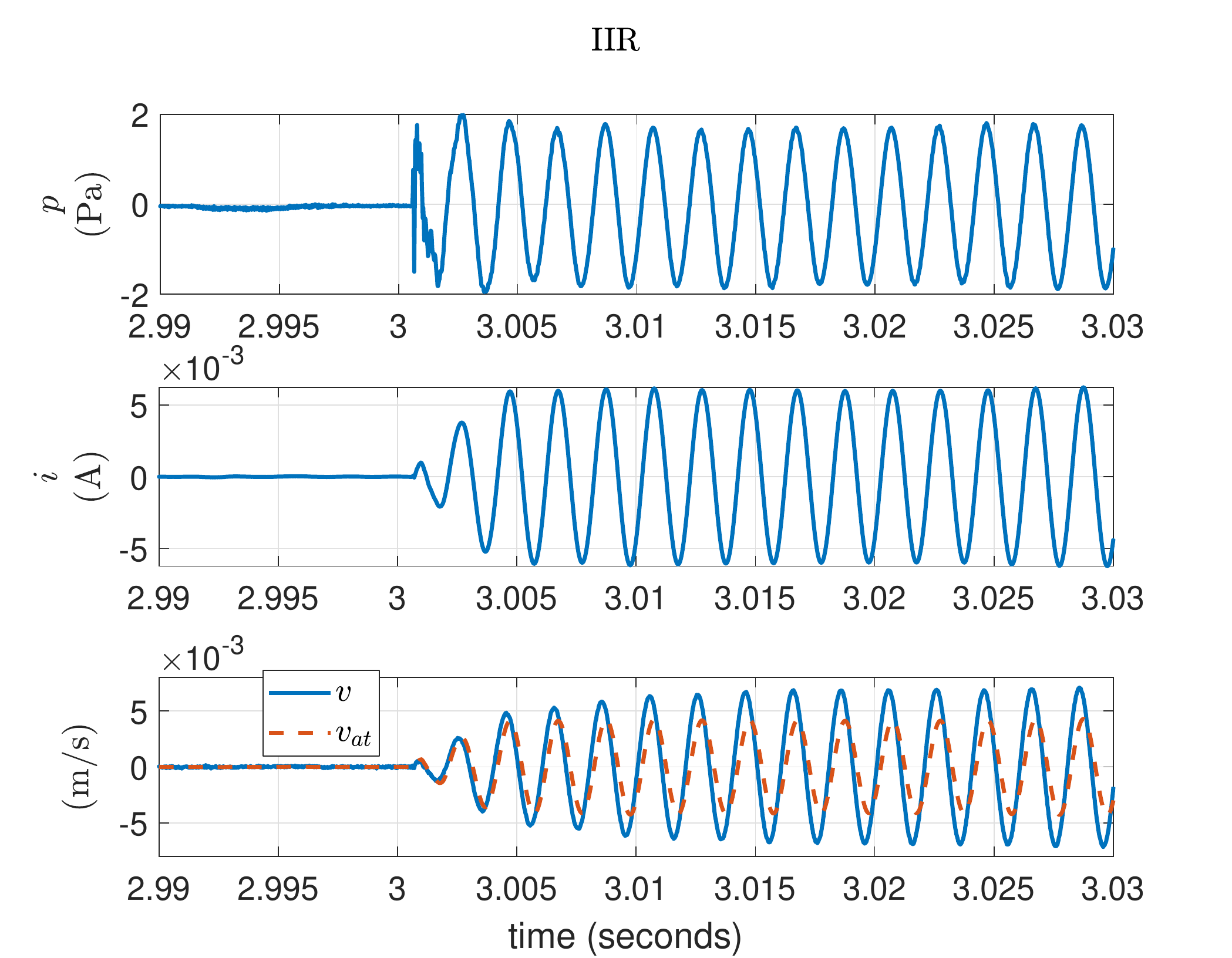}
		\centering
		\caption{}
		\label{fig:IIR_TrigExc}
	\end{subfigure}
	\hspace{1 cm}
	\begin{subfigure}[ht!]{0.45\textwidth}
		\centering
		\includegraphics[width=\textwidth]{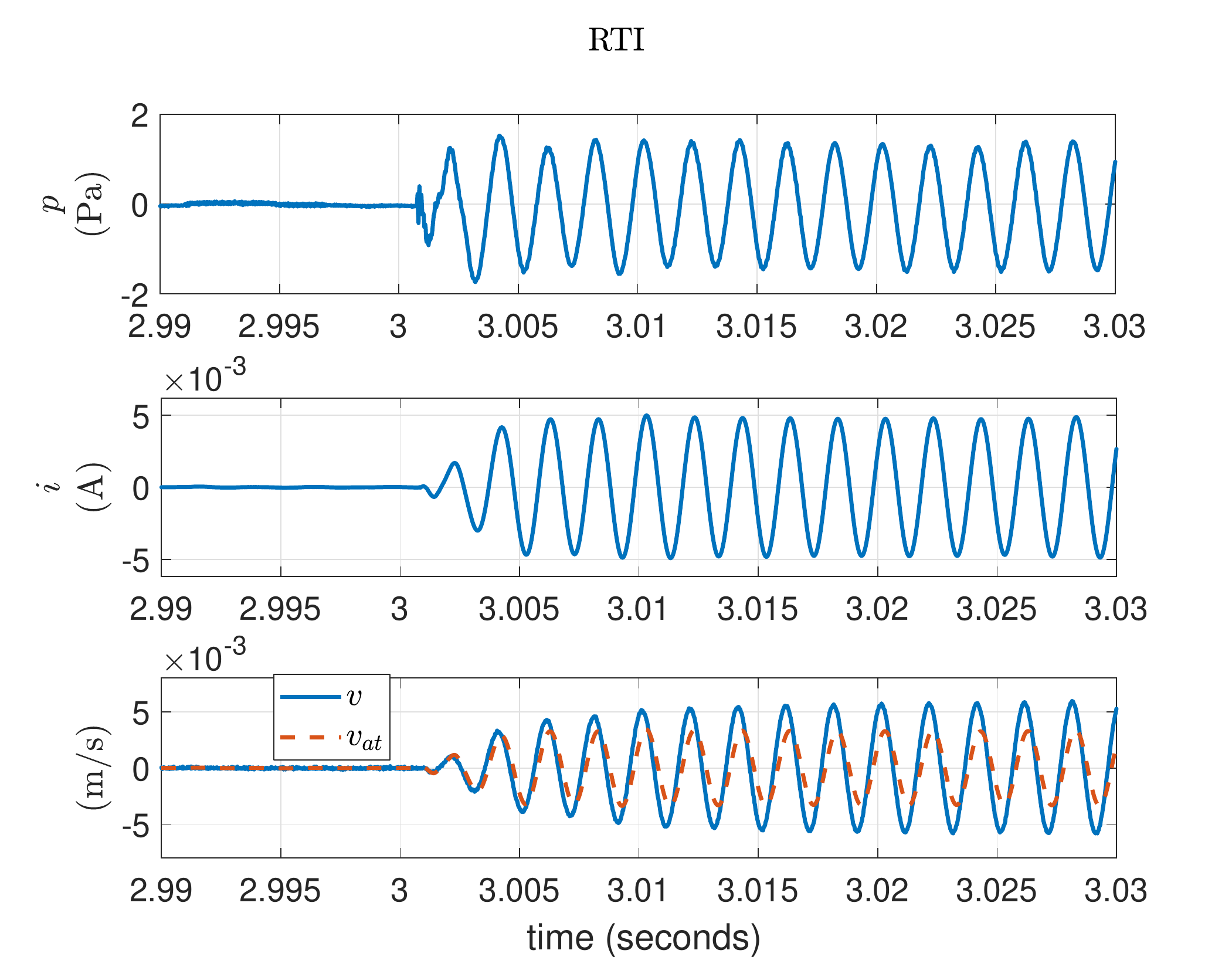}
		\caption{}
		\label{fig:RTI_TrigExc}
	\end{subfigure}
	\caption{Time histories of pressure $p(t)$, electrical current $i(t)$ and target and measured velocities ($v_{at}$ and $v(t)$ respectively) on the speaker diaphragm, when an external sound source emitting a pure sine at 500 Hz, is activated after $t=3$ seconds. A linear SDOF dynamics of the EA with $\mu_M=\mu_K=1$ and $R_{at}=\rho_0c_0$ is targeted either by the IIR \textbf{(a)} or by the RTI \textbf{(b)} implementation strategy.}
	\label{fig:TrigExc_IIRvsRTI}
\end{figure}

First, we prove the equivalence of the RTI of the EA with respect to the classical transfer-function-based IIR algorithm to target linear mechano-acoustical dynamics. The target dynamics is a linear SDOF operator relating pressure $p$ and velocity $v$, as written in Eq. \eqref{eq:target linear dynamics}.

\begin{equation}
	\label{eq:target linear dynamics}
	\begin{split}
		M_{at}\ddot{u}_{at}(t) + R_{at}\dot{u}_{at}(t) + K_{at}u_{at}(t) = p(t),
	\end{split}
\end{equation}

with target mass and stiffness which can be written in terms of the open-circuit values: $M_{at}=\mu_M M_{a0}$, $K_{at}=\mu_K K_{a0}$. The target resistance $R_{at}$ is instead defined in terms of the characteristic impedance of air $\rho_0c_0$.\\
In order to check the reliability of the RTI control strategy in following the linear target dynamics in a transient evolution, a first test has been conducted by triggering an external sound source, emitting a pure sine at 500 Hz, after 3 seconds. Figure \ref{fig:TrigExc_IIRvsRTI} shows the time histories of pressure, electrical current $i(t)$ and target and measured velocities ($v_{at}$ and $v(t)$ respectively) on the speaker diaphragm.
The target SDOF parameters are chosen as $\mu_M=\mu_K=1$ and $R_{at}=\rho_0c_0$. The results obtained by the IIR implementation are reported in Figure \ref{fig:IIR_TrigExc}, while the RTI outcomes are given in Figure \ref{fig:RTI_TrigExc}. Notice how the RTI technique allows to immediately follow the target velocity in the same way as the classical IIR. The dephasing between $v$ and $v_{at}$ depends upon the inevitable time delay in the digital control implementation of the controller, as described in \cite{DeBono2021a}, which seems to be unaffected by the control algorithm employed.

\begin{figure}[ht!]
	\centering
	\includegraphics[width=0.7\textwidth]{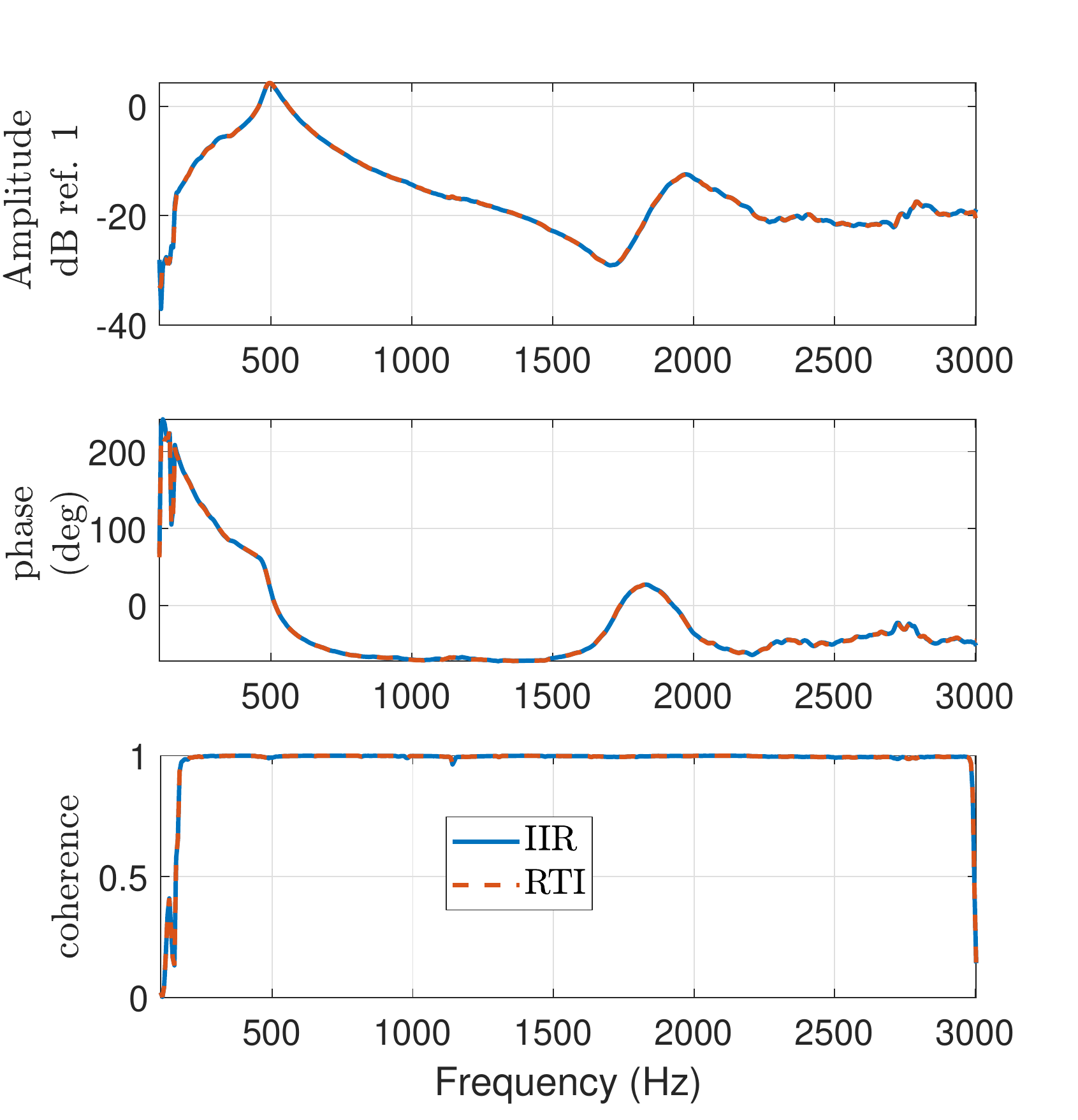}
	\caption{Mobility obtained by targeting a linear SDOF dynamics of the EA with $\mu_M=\mu_K=1$ and $R_{at}=\rho_0c_0$ by either the IIR (solid blue line) or the RTI (dashed red line).}
	\label{fig:mobility_IIRvsRTI_1muMmuK_1RatRho0c0}
\end{figure}

In Figure \ref{fig:mobility_IIRvsRTI_1muMmuK_1RatRho0c0}, the normalized mobility transfer function $\eta(\mathrm{j}\omega)=\rho_0c_0v(\mathrm{j}\omega)/p(\mathrm{j}\omega)$ is plotted in amplitude and phase, along with the coherence for $\mu_M=\mu_K=1$ and $R_{at}=\rho_0c_0$ in the frequency range 100-3000 Hz, for both IIR and RTI techniques. Observe that the two control strategy are totally equivalent. In Appendix \ref{app:IIRvsRTI tunable linear dynamics}, the target parameters of Eq. \ref{eq:target linear dynamics} are modified in order to assess the equivalence between the IIR and the RTI approaches in tuning the linear target dynamics parameters.\\
Once assessed the reliability of the RTI algorithm in the linear case, let us verify its potential to enforce nonlinear mechano-acoustical dynamics on the EA.

\subsection{Nonlinear target dynamics}\label{sec:nonlinear target dynamics}

The nonlinear mechano-acoustical dynamics targeted is the typical Duffing resonator, with cubic stiffness added to the linear one, as reported in Eq. \eqref{eq:target nonlinear dynamics}. The cubic term is multiplied by a factor called $\beta_{NL}$, which has units of $\mathrm{m^{-2}}$.

\begin{equation}
	\label{eq:target nonlinear dynamics}
	\begin{split}
		M_{at}\ddot{u}_{at}(t) + R_{at}\dot{u}_{at}(t) + K_{at}\biggr(u_{at}(t) + \beta_{NL}u^3_{at}(t)\biggr) = p(t).
	\end{split}
\end{equation}

The target mass and stiffness are defined in terms of the corresponding open-circuit values, and the target resistance expressed in terms of the characteristic impedance of air, as in the linear case (see Section \ref{sec:linear target dynamics}).

\begin{figure}
	\centering
	\includegraphics[width=.9\textwidth]{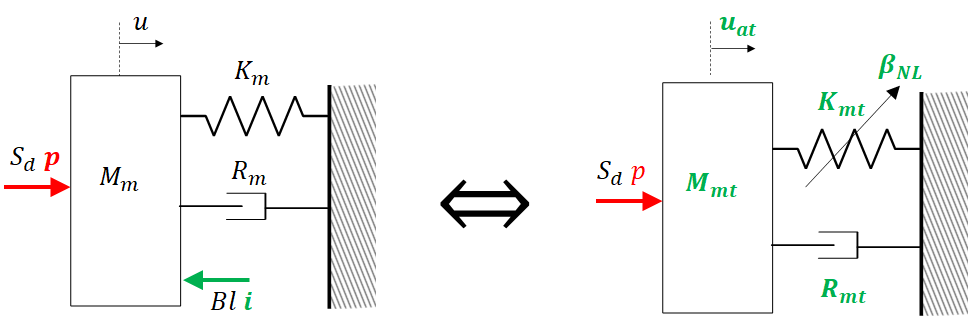}
	\caption{Sketches of the SDOF model of the EA (on the left) and of the target SDOF Duffing resonator (on the right).}
	\label{fig:EA_equivalence_NLtarget}
\end{figure}

Figure \ref{fig:EA_equivalence_NLtarget} shows the SDOF model (on the left) of the EA employed by the model inversion strategy described in Section \ref{sec:conception}, to target a SDOF Duffing resonator sketched on the right.\\

\begin{figure}[ht!]
	\centering
	\includegraphics[width=0.8\textwidth]{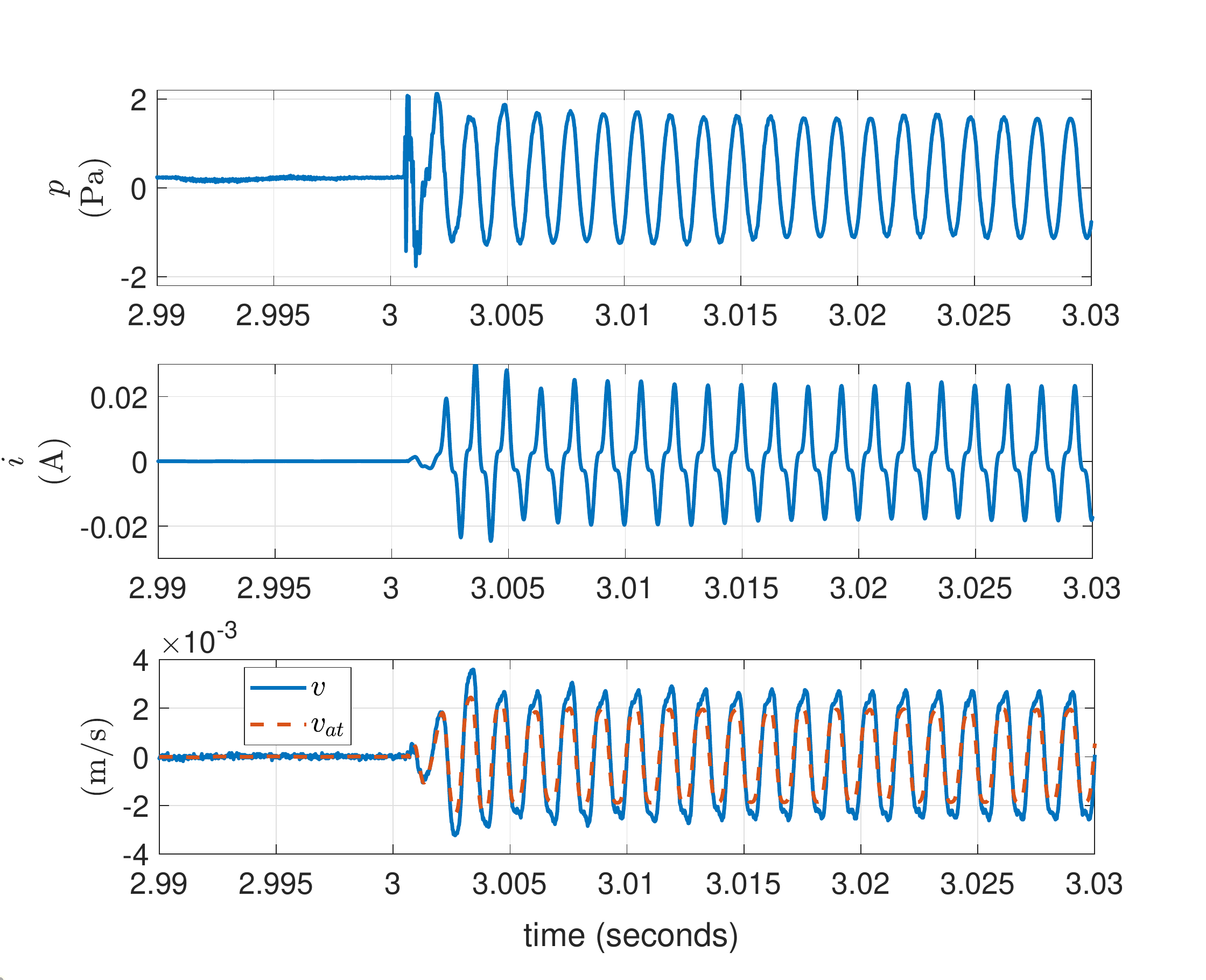}
	\caption{Time histories of pressure $p(t)$, electrical current $i(t)$, target and measured velocities ($v_{at}$ and $v(t)$ respectively) on the speaker diaphragm. An external sound source emitting a pure sine at 700 Hz, is activated after $t=3$ seconds. The linear target SDOF dynamics of the EA has $\mu_M=\mu_K=1$, $R_{at}=\rho_0c_0$ and $\beta_{NL} = 1\times10^{13}$ ($\mathrm{m^{-2}}$).}
	\label{fig:RTI_TrigExc_NonLin}
\end{figure}

\begin{figure}[ht!]
	\centering
	\begin{subfigure}[ht!]{0.45\textwidth}
		\centering
		\includegraphics[width=\textwidth]{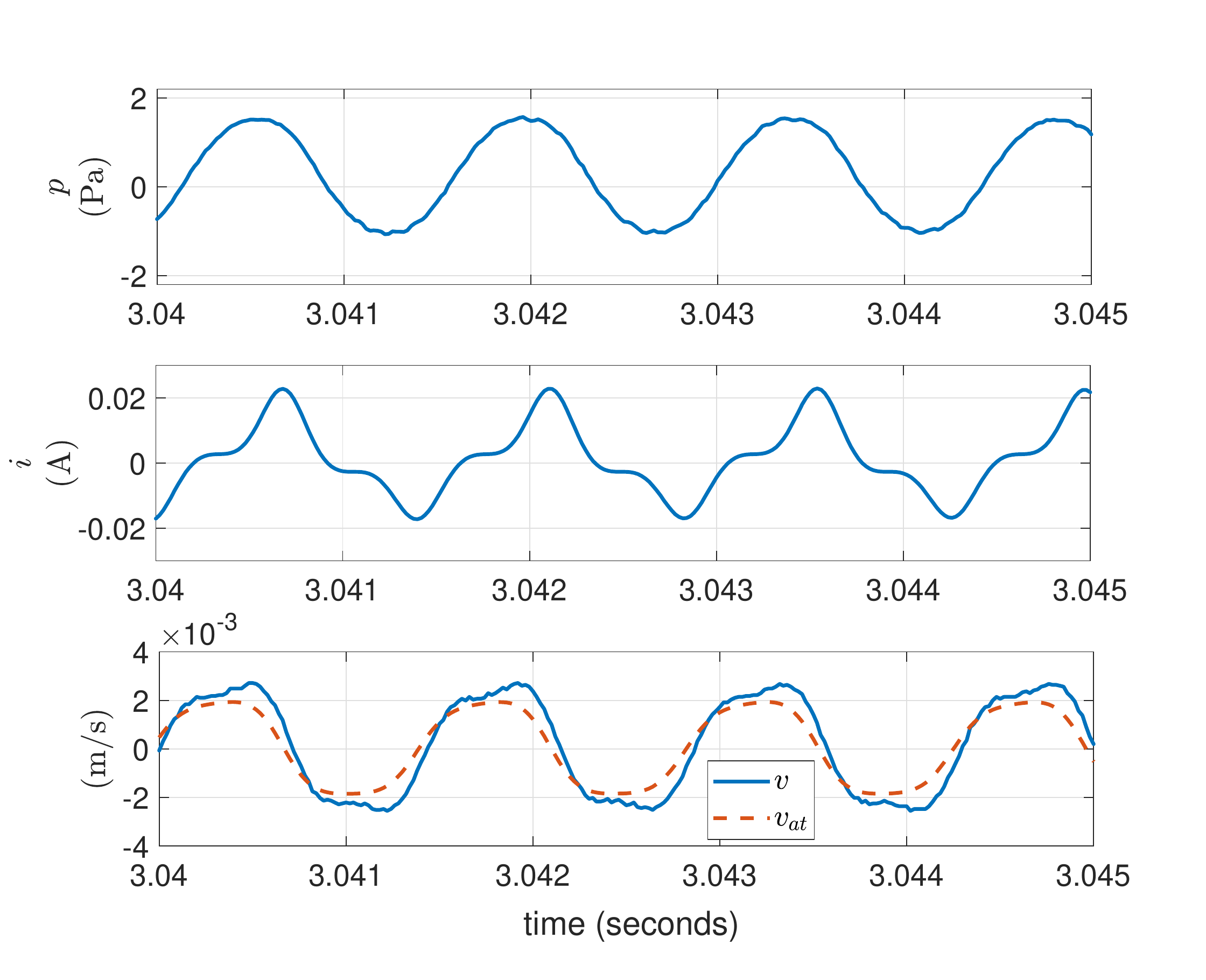}
		\centering
		\caption{}
		\label{fig:RTI_TrigExc_NonLin_ZoomStationary}
	\end{subfigure}
	\hspace{1 cm}
	\begin{subfigure}[ht!]{0.45\textwidth}
		\centering
		\includegraphics[width=\textwidth]{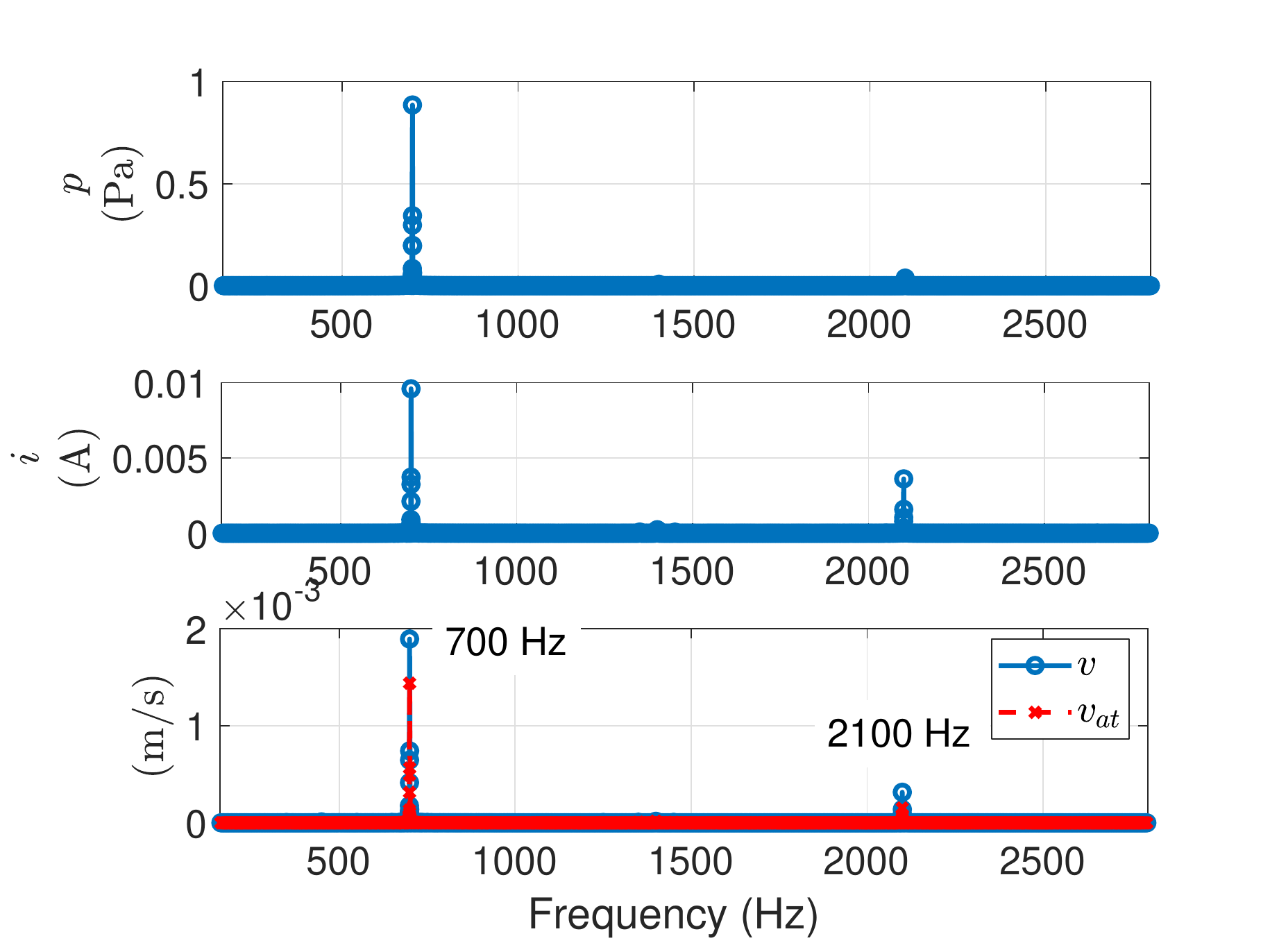}
		\caption{}
		\label{fig:RTI_TrigExc_NonLin_FFT}
	\end{subfigure}
	\caption{Zoom \textbf{(a)} of Figure \ref{fig:RTI_TrigExc_NonLin} in the stationary regime, and its FFT \textbf{(b)}.}
	\label{fig:RTI_TrigExc_NonLin_Stationary}
\end{figure}

As for the linear case, the performance of the RTI is first proved in the transient regime. Hence, an external sound source emitting a pure sine at 700 Hz is triggered after 3 seconds, and the response of the EA is measured in terms of pressure, electrical current and velocity. Figure \ref{fig:RTI_TrigExc_NonLin} shows that the model-inversion technique in the RTI algorithm is capable to follow the target velocity with the same approximation as for the linear case. Figure \ref{fig:RTI_TrigExc_NonLin_ZoomStationary} displays a zoom in the stationary regime of the signals. The electrical current clearly shows the presence of harmonics, which bring about the multi-harmonic response of the EA diaphragm as it appears in the plot of $v$. Figure \ref{fig:RTI_TrigExc_NonLin_FFT} shows the Fast-Fourier-Transform of the time signals. The system responds with a secondary resonance (related to the free oscillations) at 3 times the excitation frequency as common of Duffing resonators. 

\begin{figure}[ht!]
	\centering
	\includegraphics[width=0.8\textwidth]{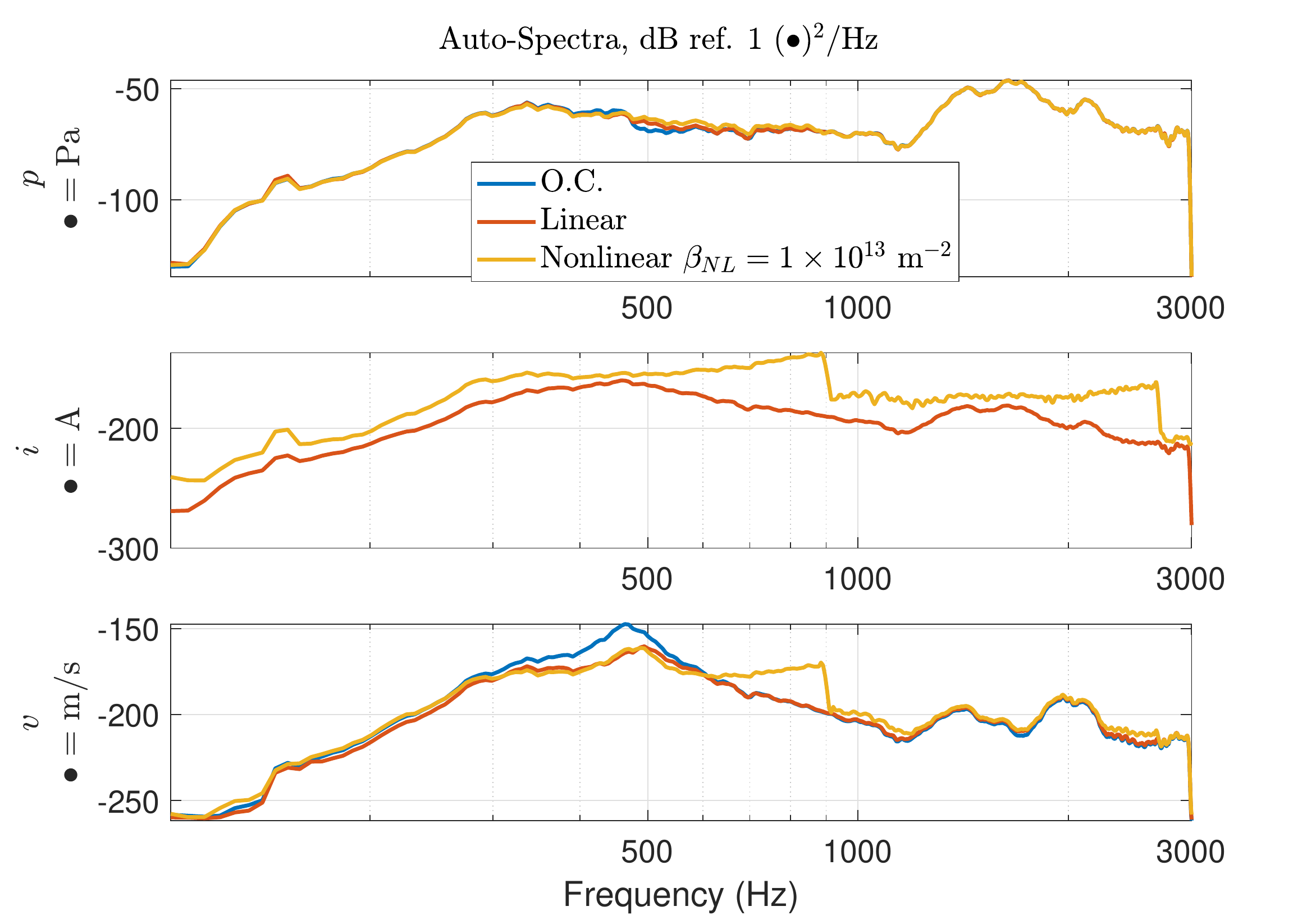}
	\caption{Auto-Spectra of pressure and velocity between 100 and 3000 Hz, in case of open circuit (O.C.) loudspeaker, Linear (with $\mu_M=\mu_K=1$ and $R_{at}=\rho_0c_0$) and Duffing (with $\beta_{NL}=1\times10^{13}$ $\mathrm{m^{-2}}$) target dynamics.}
	\label{fig:AutoSpectra_P_V_OCvsLINvsNONLIN}
\end{figure}

In Figure \ref{fig:AutoSpectra_P_V_OCvsLINvsNONLIN} the auto-spectra of pressure ($P$) on the speaker diaphragm, of electrical current ($I$) and of velocity ($V$) are plotted, comparing the Open-Circuit (O.C.) performance to the cases of control with linear or Duffing SDOF target dynamics. The autospectra are retrieved by a swept sine excitation with frequency increasing from 100 and 3000 Hz in 10 seconds. The ``jump'' phenomenon due to the cubic function of $u$ in the restoring force of the spring, typical of the Duffing resonator, clearly appears at about 890 Hz. Observe also that the dynamics at the original (O.C.) resonance of the EA is not totally cancelled out by the model-inversion control. This is also due to the time delay as described in \cite{DeBono2021}. Notice that the pressure autospectra in Figure \ref{fig:AutoSpectra_P_V_OCvsLINvsNONLIN} are almost the same for each configuration, demonstrating that the nonlinear behaviour is achieved thanks to the control and is not naturally triggered in the speaker by high excitation levels. The target SDOF resonator parameters are $M_{at}=M_{a0}$, $K_{at}=K_{a0}$ and $R_{at}=\rho_0c_0$, while the cubic term in the Duffing case is multiplied by a $\beta_{NL}=1\times10^{13}$ ($\mathrm{m^{-2}}$). Such a huge value for $\beta_{NL}=1\times10^{13}$ $\mathrm{m^{-2}}$ is needed in order to enforce a cubic behaviour of electrical current at moderate excitation levels, as showed by the autospectrum of $i$. Through the Lorentz force, the cubic behaviour is transferred to the mechanical dynamics. As higher $\beta_{NL}$ is, as lower excitation levels are necessary to trigger the same nonlinear phenomena.\\
Notice also the jump phenomenon of the superharmonic resonance at about 2670 Hz, which is exactly 3 times the primary nonlinear resonance frequency relative to the \emph{fundamental} harmonics. The scattered superharmonic field, at 3 times the excitation frequency indeed, also produces a resonance (the so-called \emph{superharmonic resonance} \cite{nayfeh1980nonlinear}), with bending and jump of the frequency response. This jump is evident in the electrical current plot, and slightly in the velocity spectrum. 

\begin{figure}[ht!]
	\centering
	\includegraphics[width=0.8\textwidth]{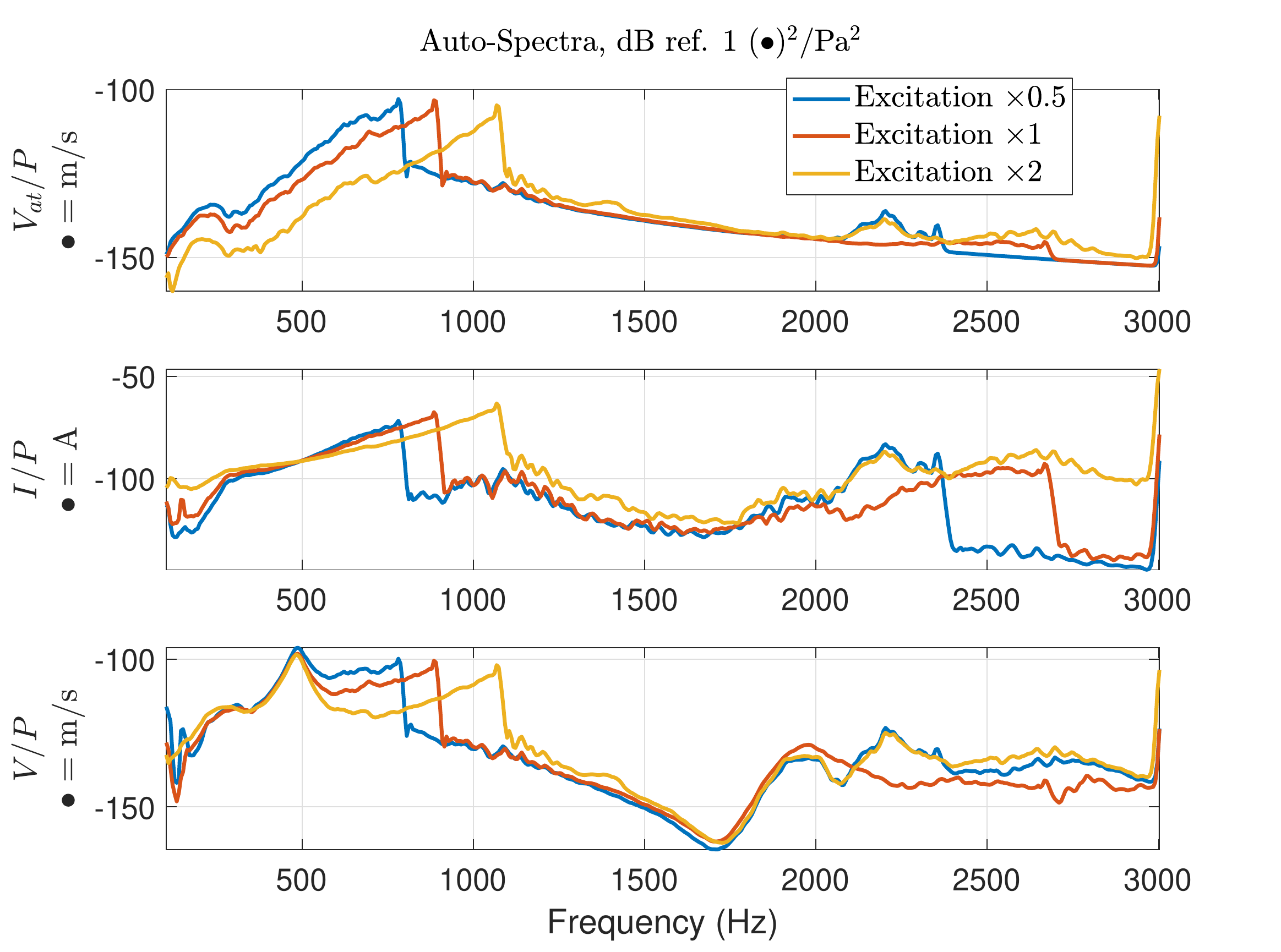}
	\caption{Auto-Spectra of target velocity, electrical current and actual velocity, divided by the autospectrum of pressure, for varying excitation amplitude. The terms of the target dynamics are fixed as $\mu_M=\mu_K=1$, $R_{at}=\rho_0c_0$ and $\beta_{NL}=1\times10^{13}$ $m^{-2}$.}
	\label{fig:AutoSpectraOverP_varying_AmplExc}
\end{figure}

Since a nonlinear dynamics is targeted, multi-harmonic scattering is expected by our EA, and the transfer functions typically employed to characterize linear responses, lose their significance. In the attempt to present figures which are analogous to transfer functions, in the following we choose to report the autospectra of each variable of interest, divided by the autospectrum of pressure. Remind that the multi-harmonic scattered field entails also a multi-harmonic excitation field, as the \emph{fundamental} harmonic waves are not physically uncoupled by the \emph{non-fundamental} ones, as it would be in a perfectly open field environment. Hence, both the autospectra and their ratios, do not distinguish between fundamental and higher harmonics. To retrieve the fundamental and higher harmonics separately, a dedicated experimental setup should be conceived, which will be the subject of another contribution of this article series, where both \emph{acoustical passivity} \cite{DeBono2021} and absorptive performance of the achieved Duffing resonator dynamics will be investigated.\\  
As a typical feature of nonlinear systems is the dependence of the frequency response upon the excitation amplitude, Figure \ref{fig:AutoSpectraOverP_varying_AmplExc} shows the effect of varying the external acoustic source signal to half and twice the one employed so far. As the excitation increases, the jump moves toward higher frequencies, as typical of Duffing resonators. Observe that the jump is present also when the external acoustic excitation is halved. As expected, the effect of increasing the external excitation is similar to the one of augmenting $\beta_{NL}$, as both enhance the nonlinear Lorentz force. 

\begin{figure}[ht!]
	\centering
	\centering
	\includegraphics[width=0.8\textwidth]{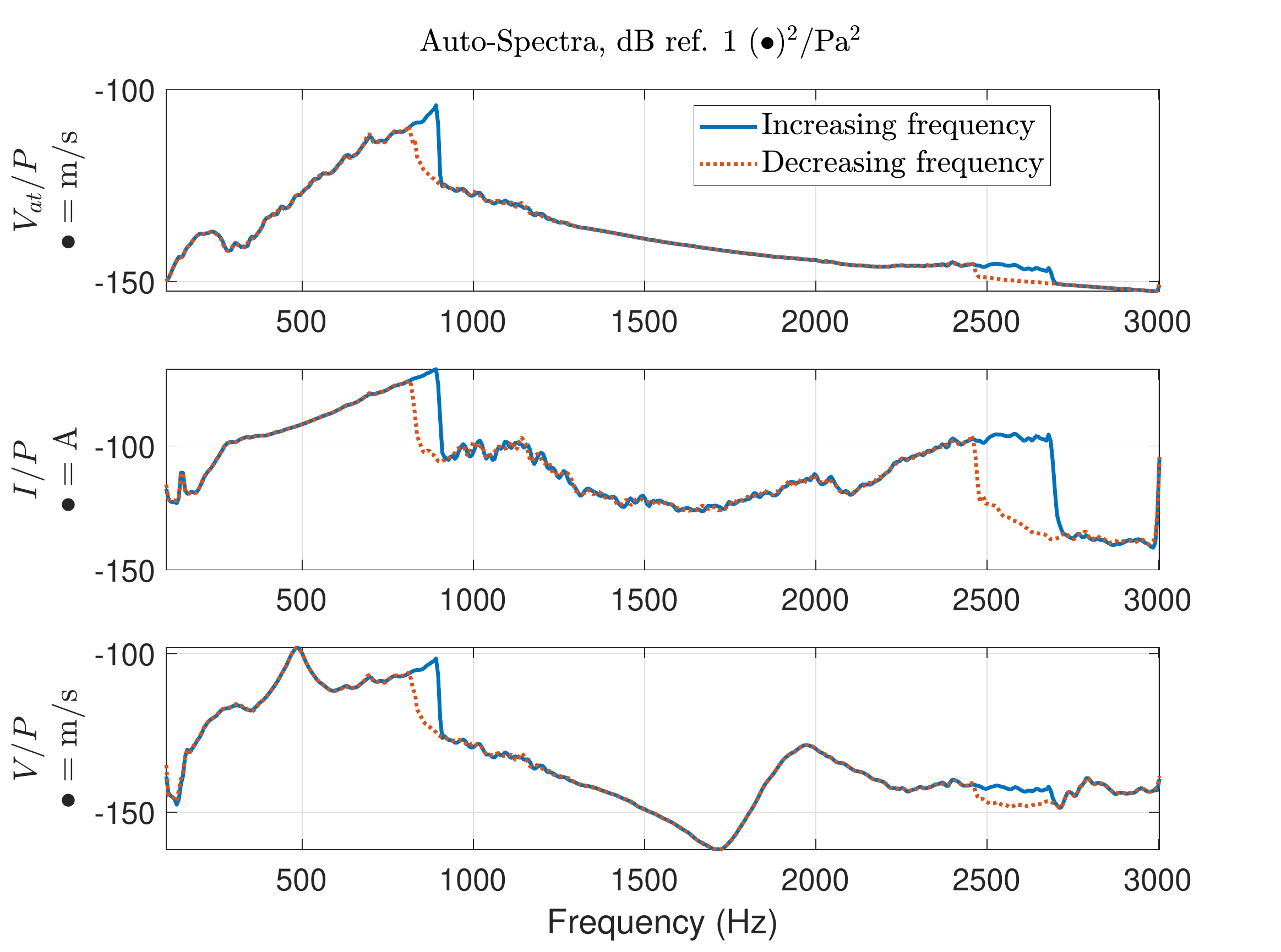}
	\caption{Auto-Spectra of target velocity, electrical current and actual velocity, divided by the autospectrum of pressure, for \emph{forward} or \emph{backard} linear variation of frequency in the swept sine. The terms of the target dynamics are fixed as $\mu_M=\mu_K=1$, $R_{at}=\rho_0c_0$ and $\beta_{NL}=1\times10^{13}$ $m^{-2}$.}
	\label{fig:AutoSpectraOverP_FORWARDvsBACKWARD}
\end{figure}

Another characteristic of the nonlinear systems is the dependence upon the initial conditions. In particular, a cubic stiffness term realizes two possible stable responses in certain frequency ranges. The initial conditions determines which of the possible responses actually develops. Figure \ref{fig:AutoSpectraOverP_FORWARDvsBACKWARD} shows the responses in a \emph{quasi-stationary} regime realized by a swept sine excitation of 30 seconds with frequency varying between 100 and 3000 Hz in the \emph{increasing} or \emph{decreasing} sense. By varying the frequency in decreasing sense, the jumps shift toward lower frequencies.

\begin{figure}[ht!]
	\centering
	\includegraphics[width=0.8\textwidth]{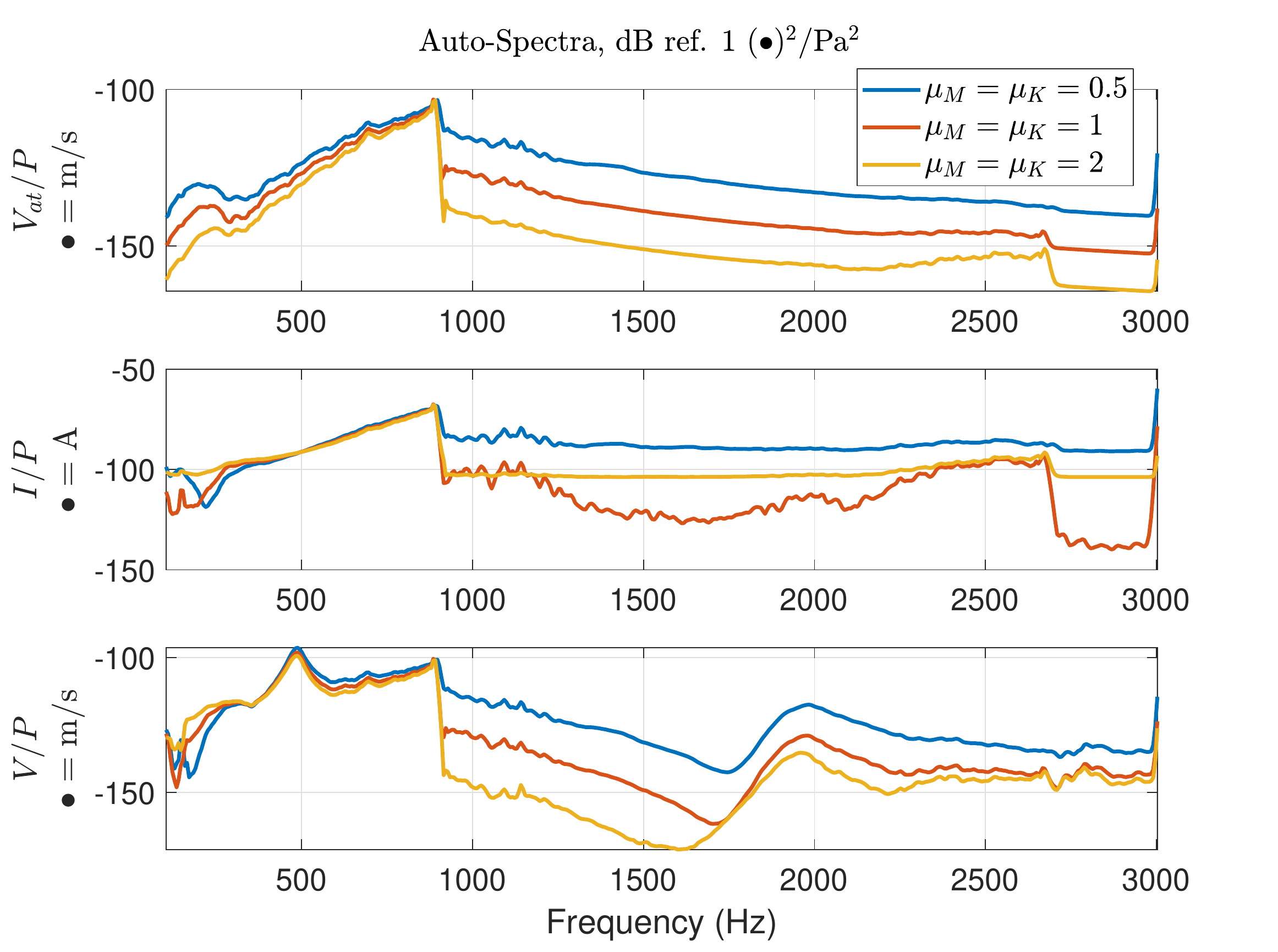}
	\caption{Auto-Spectra of target velocity, electrical current and actual velocity, divided by the autospectrum of pressure, for varying target mass and stiffness proportionally. The target resistance is $\rho_0c_0$}
	\label{fig:AutoSpectraOverP_varying_muMK}
\end{figure}

In the following figures, we demonstrate the ability to tune each of the term in the target Duffing resonator dynamics, by our RTI control. Figures \ref{fig:AutoSpectraOverP_varying_muMK} shows the tunability of the Duffing resonator with hardening spring to variable target mass and linear stiffness terms proportionally. The target resistance is $R_{at}=\rho_0c_0$ and the nonlinear coefficient is $\beta_{NL}=1\times10^{13}$ $\mathrm{m^{-2}}$. As expected, increasing the reactive part of the target dynamics, tightens the resonance pick of the hardening spring frequency response, leading to \emph{deeper} jumps at the nonlinear resonance. Looking at the plots of the actual velocity, one can notice the presence of a residual pick at the O.C. resonance (around 500 Hz), which the model-inversion technique does not manage to fully cancel out, mainly due to time delay in the control loop, as well as to parameter incertitudes.

\begin{figure}[ht!]
	\centering
	\includegraphics[width=0.8\textwidth]{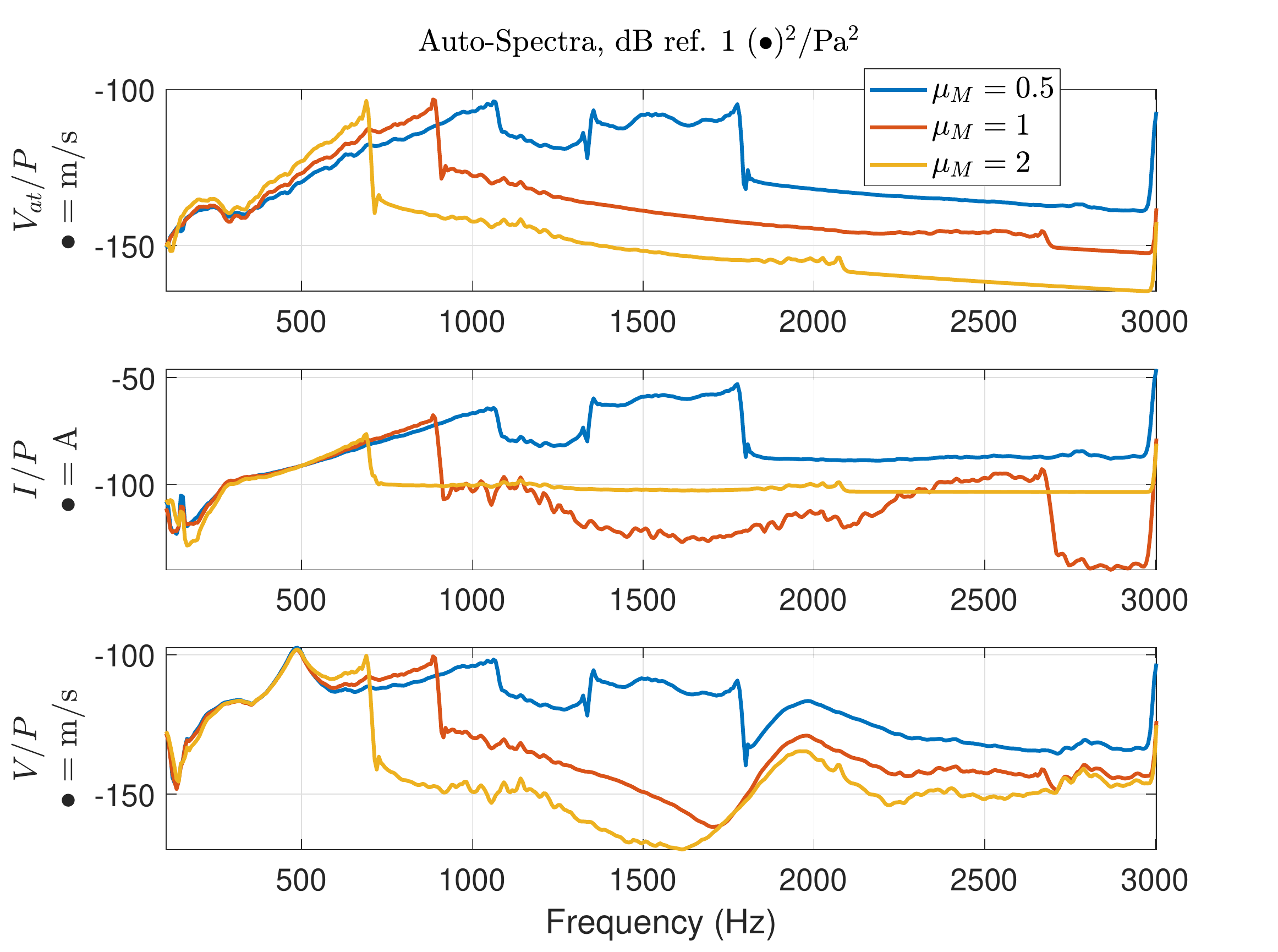}
	\caption{Auto-Spectra of target velocity, electrical current and actual velocity, divided by the autospectrum of pressure, for varying target mass. The target resistance is $\rho_0c_0$ and the target stiffness factor is $\mu_K=1$.}
	\label{fig:AutoSpectraOverP_varying_muM}
\end{figure}

In Figure \ref{fig:AutoSpectraOverP_varying_muM} the target mass term is varied, showing the expected shifts of the nonlinear resonance, along with the residual O.C. resonance in the actual velocity plot. Observe that for $\mu_M=0.5$ two additional jumps appear at 1355 and 1776 Hz. The one at 1776 Hz is probably a super-harmonic resonance due to a two-term excitation \cite{nayfeh1980nonlinear}, i.e. the simultaneous presence of two harmonics exciting the system. The first harmonic can be the primary resonance ($\Omega_1=1068$ Hz), while the second one can be the subharmonic scattered field at one-third of the primary resonance ($\Omega_2=356$ Hz). Because of the afore-mentioned coupling between the scattered and incident fields on the EA, the scattered subharmonic wave is exciting back our system, and a two-frequency excitation can occur. In this case, a secondary resonance can occur at $2\Omega_1-\Omega_2=1780$ Hz, which is approximately the frequency (1776 Hz) of one unexpected jump. The jump at 1355 Hz instead, looks as corresponding to a \emph{softening spring} bending of the frequency response. In order to understand such phenomenon, it might be useful to check possible spillover effects on the nearby mode of the speaker. These unexpected phenomena will be better investigated if the incident and scattered field are separated, which is the objective of another contribution.\\

\begin{figure}[ht!]
	\centering
	\includegraphics[width=0.8\textwidth]{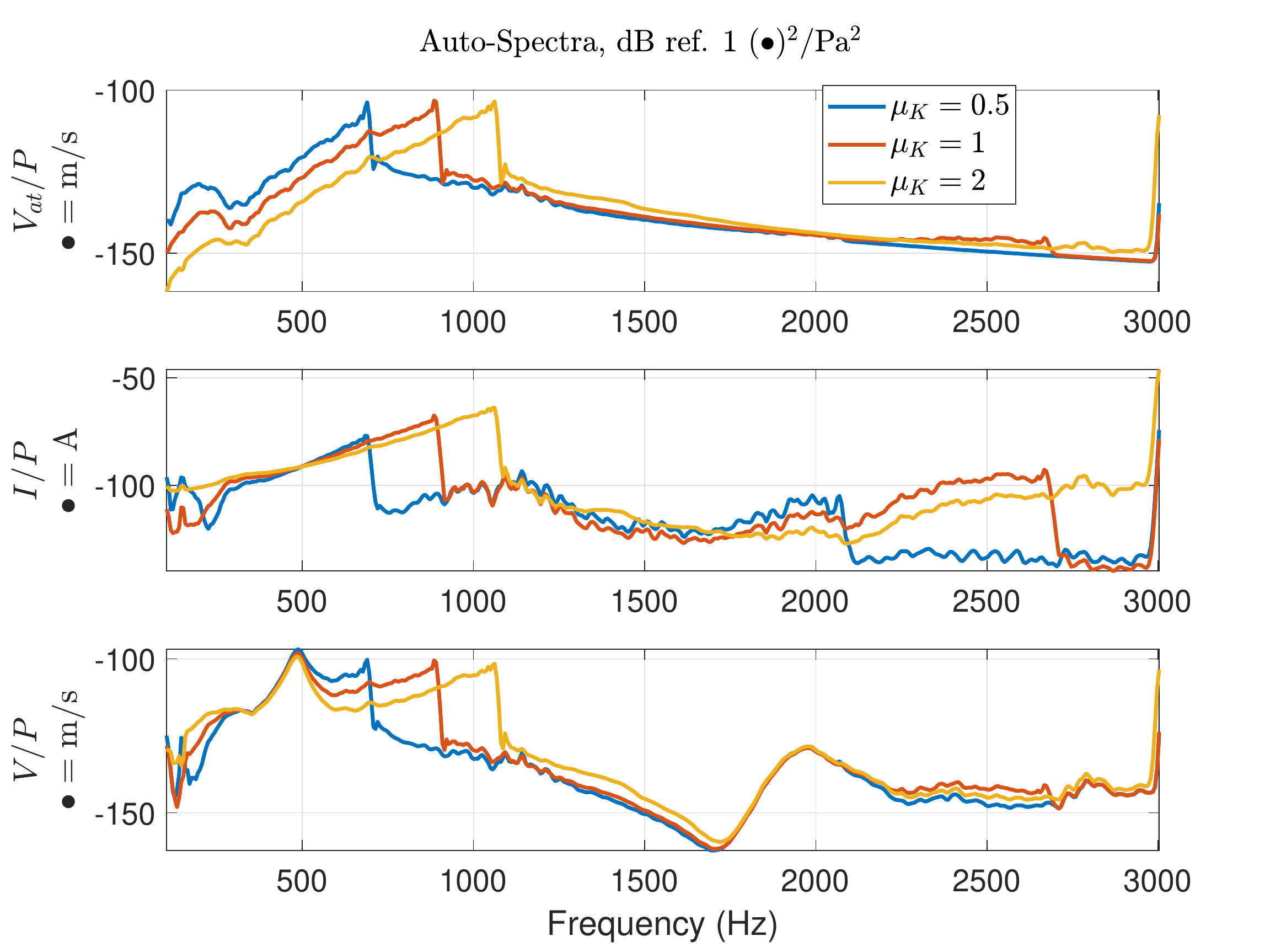}
	\caption{Auto-Spectra of target velocity, electrical current and actual velocity, divided by the autospectrum of pressure, for varying target stiffness. The target resistance is $\rho_0c_0$ and the target mass factor is $\mu_M=1$.}
	\label{fig:AutoSpectraOverP_varying_muK}
\end{figure}

Figure \ref{fig:AutoSpectraOverP_varying_muK} shows the effect of varying the linear target stiffness term. As the linear resonance is shifted toward higher frequency, same it does the nonlinear one.

\begin{figure}[ht!]
	\centering
	\includegraphics[width=0.8\textwidth]{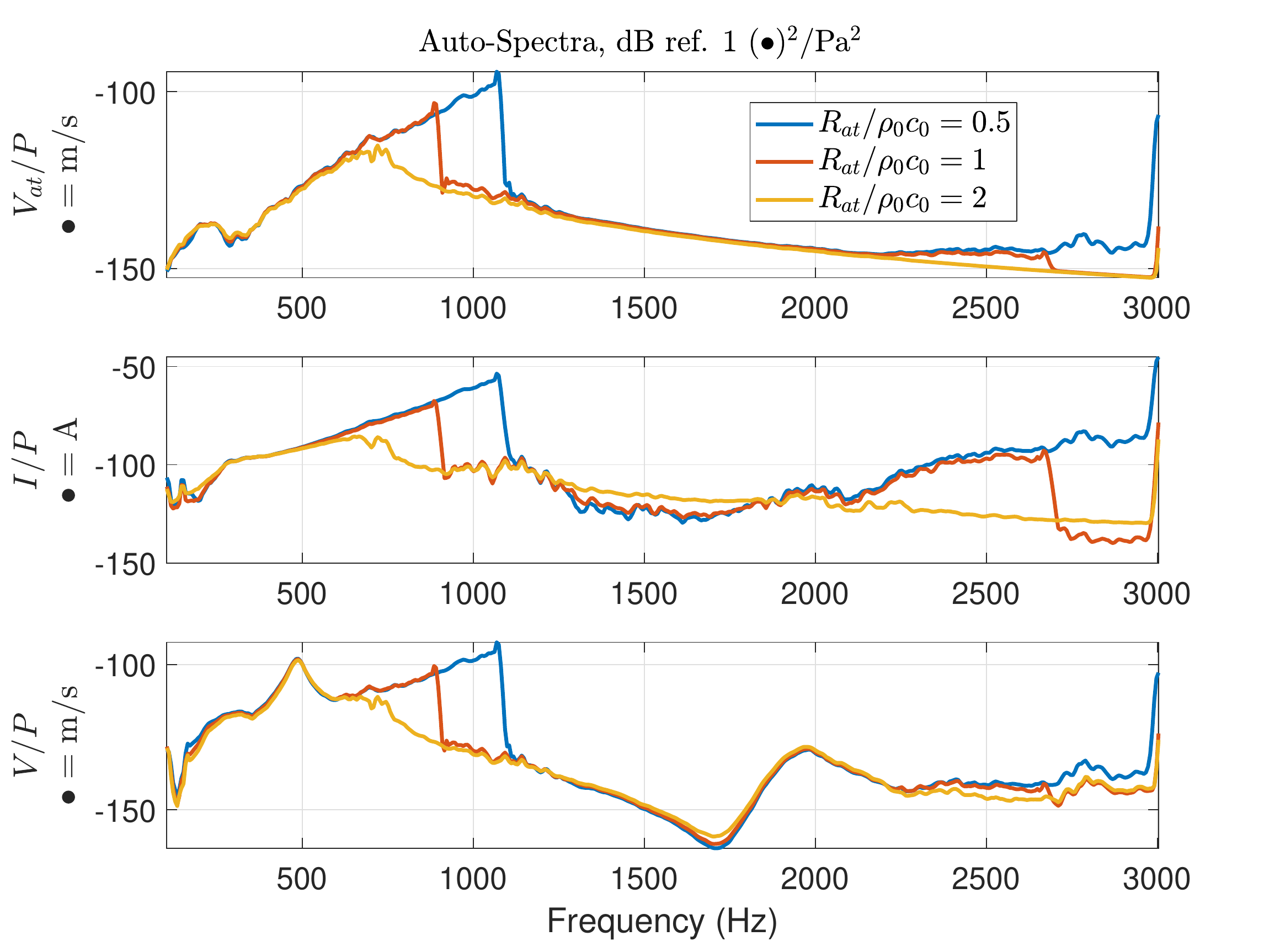}
	\caption{Auto-Spectra of target velocity, electrical current and actual velocity, divided by the autospectrum of pressure, for varying target resistance. The target mass and stiffness factors are $\mu_M=\mu_K=1$.}
	\label{fig:AutoSpectraOverP_varying_Rat}
\end{figure}

The effect of varying the target resistance $R_{at}$ is showed in Figure \ref{fig:AutoSpectraOverP_varying_Rat}. As expected, reducing the resistance sharpens and heightens the pick, leading to deeper jumps.

\begin{figure}[ht!]
	\centering
	\includegraphics[width=0.8\textwidth]{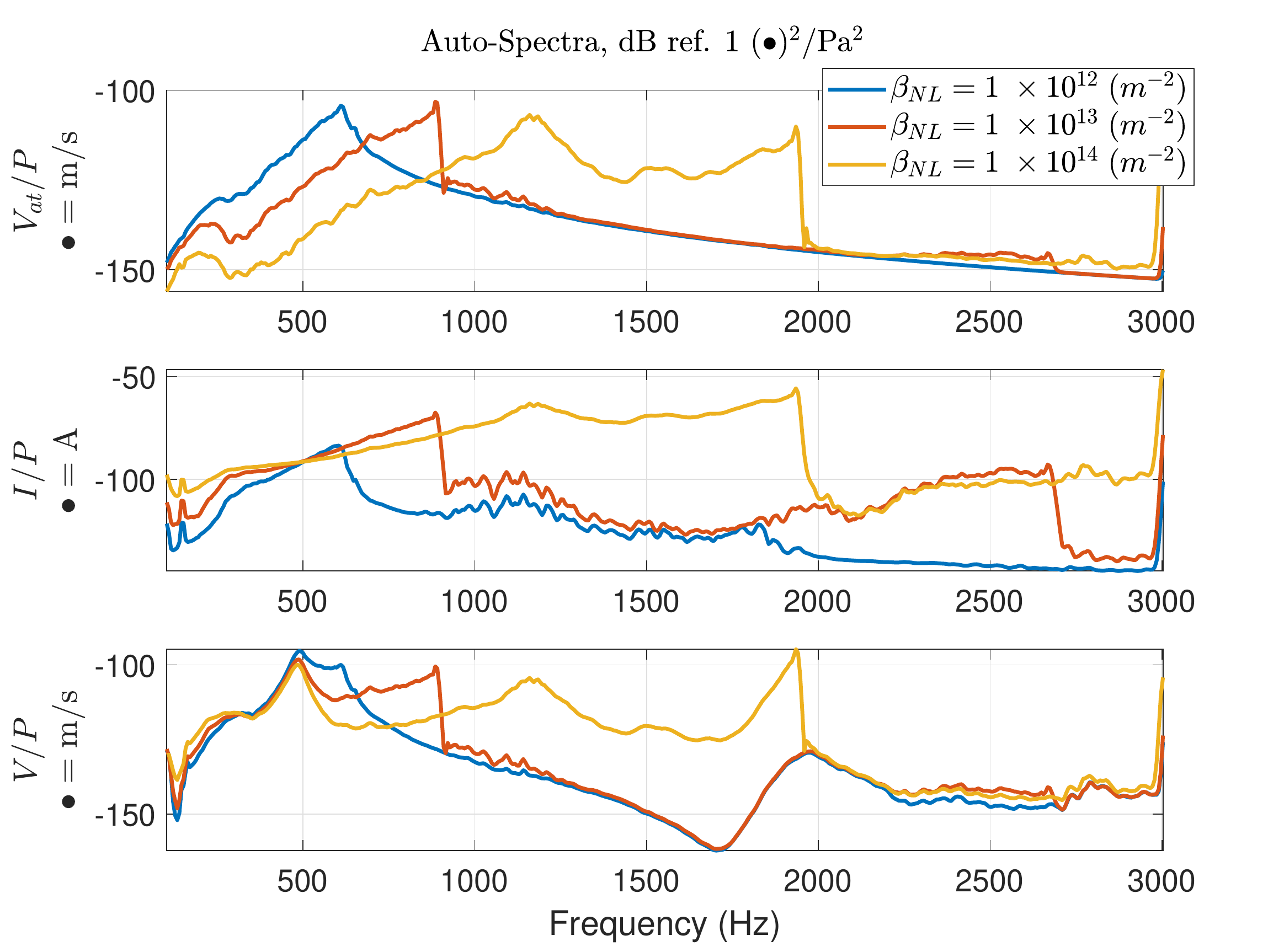}
	\caption{Auto-Spectra of target velocity, electrical current and actual velocity, divided by the autospectrum of pressure, for varying the nonlinear coefficient $\beta_{NL}$. The target mass and stiffness factors are $\mu_M=\mu_K=1$, and the target resistance is $R_{at}=\rho_0c_0$.}
	\label{fig:AutoSpectraOverP_varying_betaNL}
\end{figure}

The effect of varying $\beta_{NL}$ is displayed in Figure \ref{fig:AutoSpectraOverP_varying_betaNL}. As expected, increasing $\beta_{NL}$ moves the nonlinear resonance, and relative jump, toward higher frequencies. Observe that the velocity on the speaker diaphragm is affected by the additional mode at about 2000 Hz (which is not taken into account in the model-inversion control). 

\section{Conclusions}\label{sec:conclusions}

This paper has proposed an innovative strategy to enforce a tunable Duffing-type dynamics on the Electroacoustic Absorber, based upon the sole measurement of the pressure on the speaker diaphragm. As the measurement of the electroacoustic absorber diaphragm displacement $u(t)$ was not available, a Duffing-type response could not be achieved by simply inserting the cubic power of $u(t)$ in the controller (as proposed in \cite{guo2020improving}), and an alternative strategy had to be devised. We have here proposed a control algorithm which is based upon the real-time-integration of a (possibly nonlinear) target dynamics, and upon the model-inversion. The peculiarity of such strategy is the possibility to enforce nonlinear behaviour when the controlled system would normally behave as linear. A large literature is available in the control of nonlinear systems, but, to the authors' knowledge, no significant contribution in the \emph{enforcement} of a nonlinear equivalent behaviour has so far been proposed, except \cite{guo2020improving}. Clearly, the possibility to exploit a measurement of the motion of the speaker (such as by a microphone in the back cavity as in \cite{guo2020improving} or other) in our Real-Time-Integration control scheme, might allow to reduce the problems of residual original dynamics and the difficulties of model inversion. Both this and other limitations which typically concern any model-inversion scheme relate to the model and parameters uncertainties, might also be tackled by enlarging optimization algorithms (such as $H_{\infty}$) toward nonlinear target dynamics.\\
The experimental validation has confirmed the potentiality of such control scheme to enforce a tunable Duffing resonator dynamics with \emph{hardening}-spring behaviour on the EA diaphragm. Future contributions will analyse other possible nonlinear responses of interest, such as ones involving essential nonlinearities. The experimental campaign will continue toward the investigation of both passivity and absorptive performances of our nonlinear absorbers, by an ad-hoc conceived test-rig, as the ones proposed in \cite{boden2012one}, \cite{bellet2010experimental}. The advantage respect to the classical problem of nonlinear absorber characterization is that our device does not require high excitation amplitudes, hence the acoustic field will keep its linear character, allowing to envisage the feasibility to physically separate the incident from the multi-harmonic scattered fields.\\
Numerical and experimental analyses are foreseen to verify the Nonlinear Energy Sink capabilities of our electro-active absorber, as well as its optimization. The preliminary results presented in this contribution have indeed opened the doors toward nonlinear programmable boundaries also in acoustics, encouraging the cutting edge research in the inverse problem of nonlinear absorber design.

\FloatBarrier
\begin{appendices}
	\section{Comparing IIR and RTI in linear target dynamics tuning}\label{app:IIRvsRTI tunable linear dynamics}
	
\begin{figure}[ht!]
		\centering
		\begin{subfigure}[ht!]{0.45\textwidth}
			\centering
			\includegraphics[width=\textwidth]{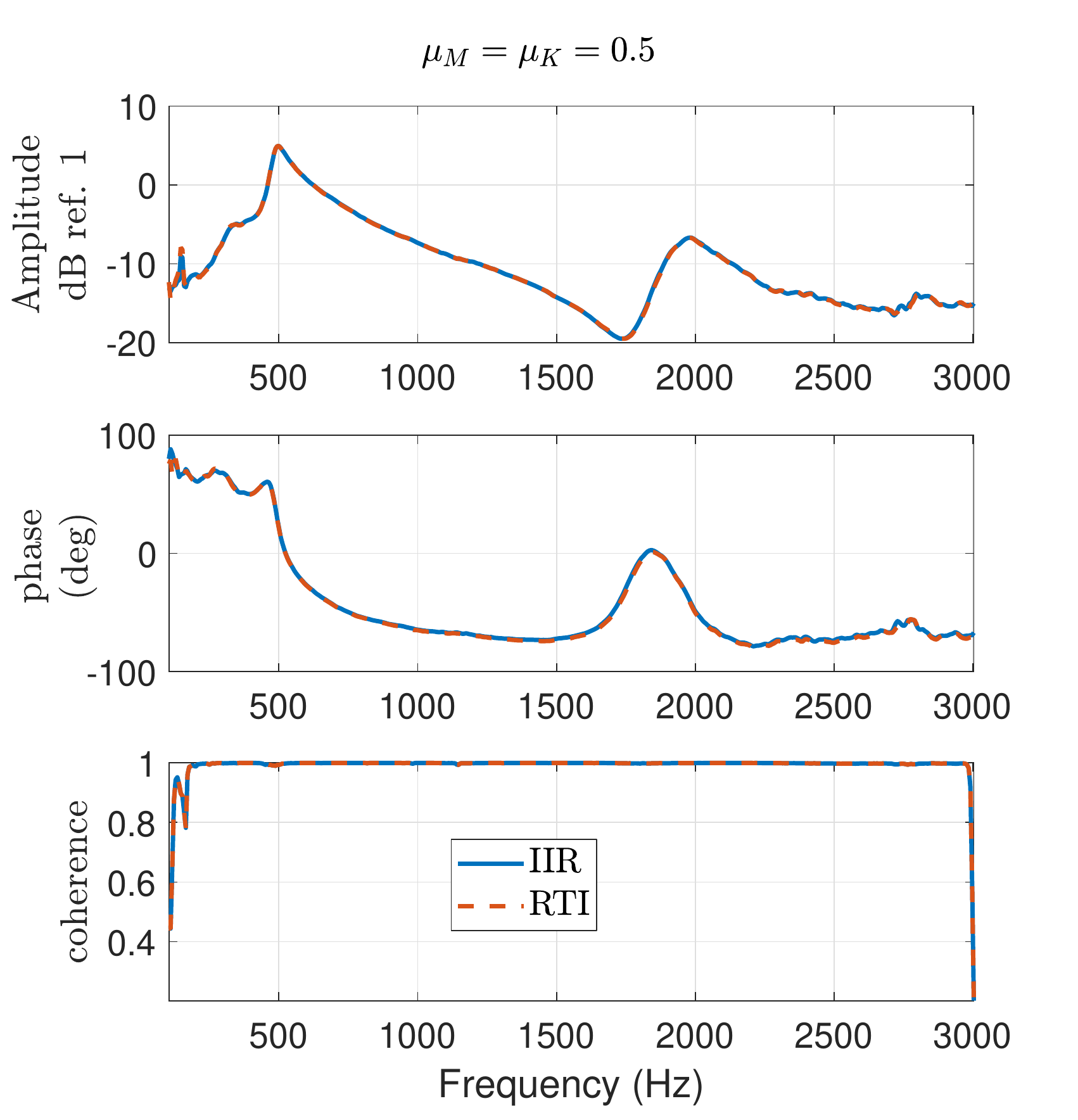}
			\centering
			\caption{}
			\label{fig:mobility_IIRvsRTI_0.5muMmuK}
		\end{subfigure}
		\hspace{0.5 cm}
		\begin{subfigure}[ht!]{0.45\textwidth}
			\centering
			\includegraphics[width=\textwidth]{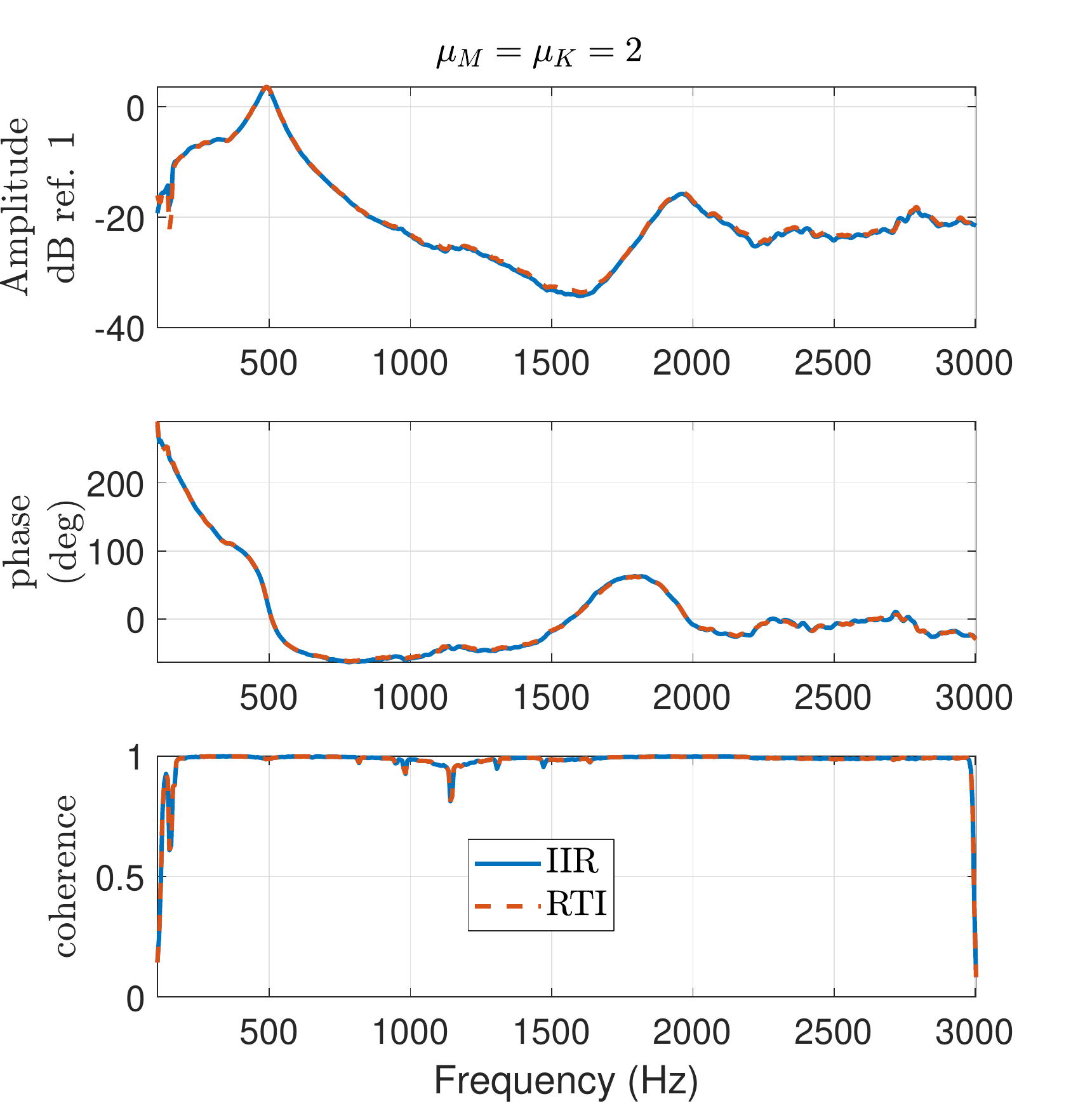}
			\caption{}
			\label{fig:mobility_IIRvsRTI_2muMmuK}
		\end{subfigure}
		\caption{Mobility obtained by targeting a linear SDOF dynamics of the EA with $\mu_M=\mu_K=0.5$ \textbf{(a)} and $\mu_M=\mu_K=2$ \textbf{(a)}, and $R_{at}=\rho_0c_0$, by either the IIR (solid blue line) or the RTI (dashed red line).}
		\label{fig:mobility_IIRvsRTI_Varying_muMmuK}
	\end{figure}

\begin{figure}[ht!]
		\centering
		\begin{subfigure}[ht!]{0.45\textwidth}
			\centering
			\includegraphics[width=\textwidth]{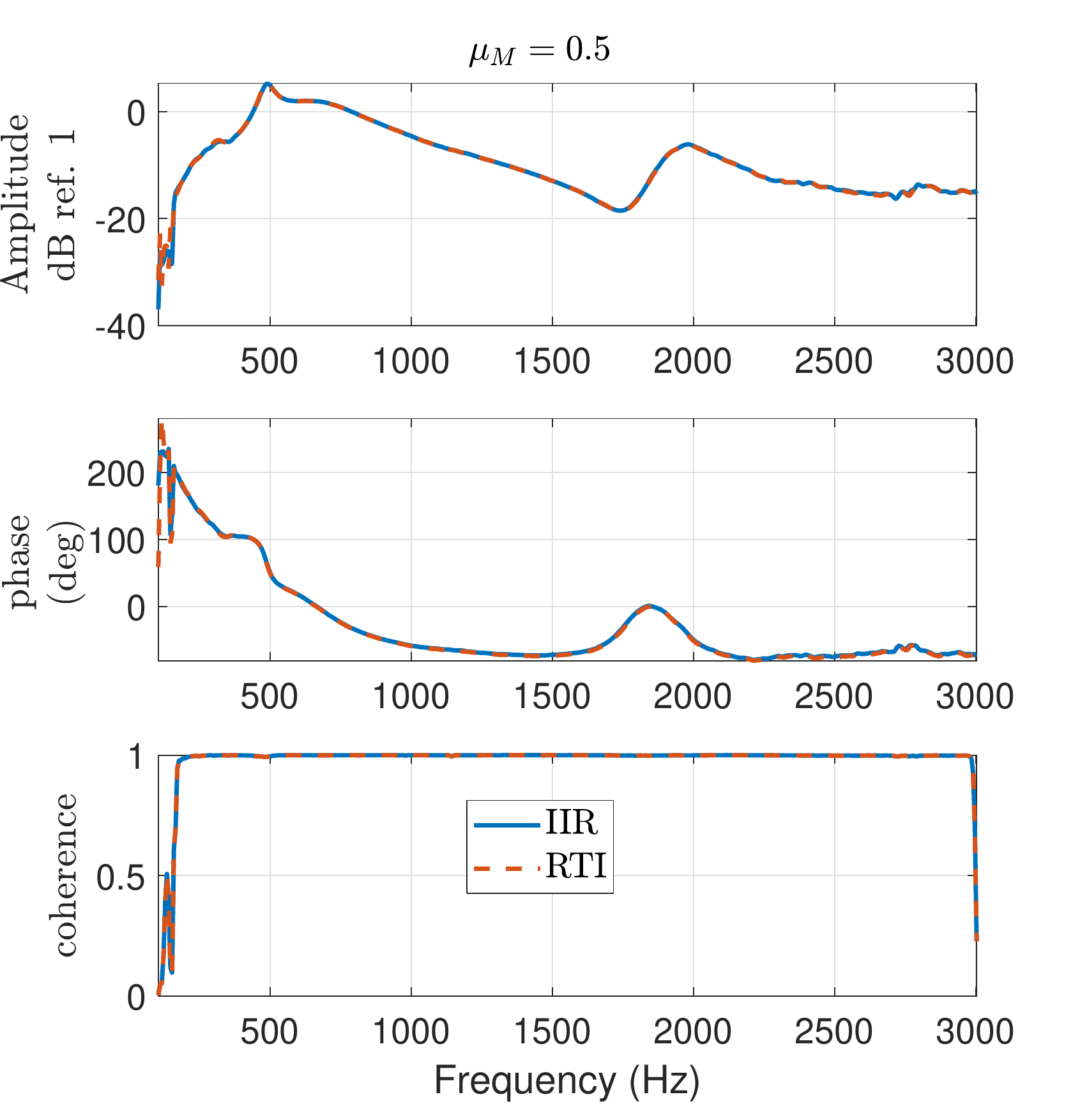}
			\centering
			\caption{}
			\label{fig:mobility_IIRvsRTI_0.5muM}
		\end{subfigure}
		\hspace{0.5 cm}
		\begin{subfigure}[ht!]{0.45\textwidth}
			\centering
			\includegraphics[width=\textwidth]{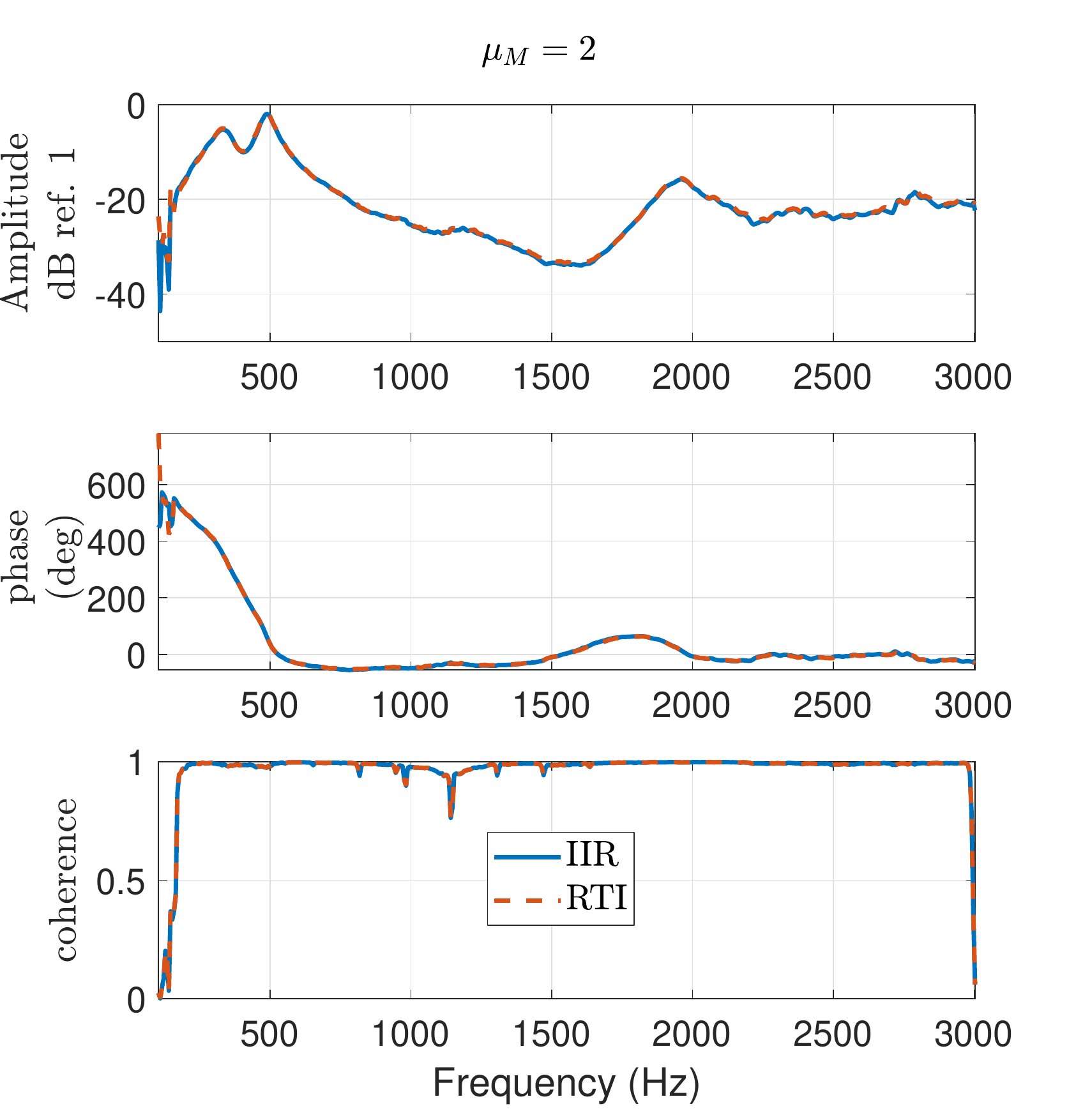}
			\caption{}
			\label{fig:mobility_IIRvsRTI_2muM}
		\end{subfigure}
		\caption{Mobility obtained by targeting a linear SDOF dynamics of the EA with $\mu_M=0.5$ \textbf{(a)} and $\mu_M=2$ \textbf{(a)}, $\mu_K=1$ and $R_{at}=\rho_0c_0$, by either the IIR (solid blue line) or the RTI (dashed red line).}
		\label{fig:mobility_IIRvsRTI_Varying_muM}
	\end{figure}

\begin{figure}[ht!]
		\centering
		\begin{subfigure}[ht!]{0.45\textwidth}
			\centering
			\includegraphics[width=\textwidth]{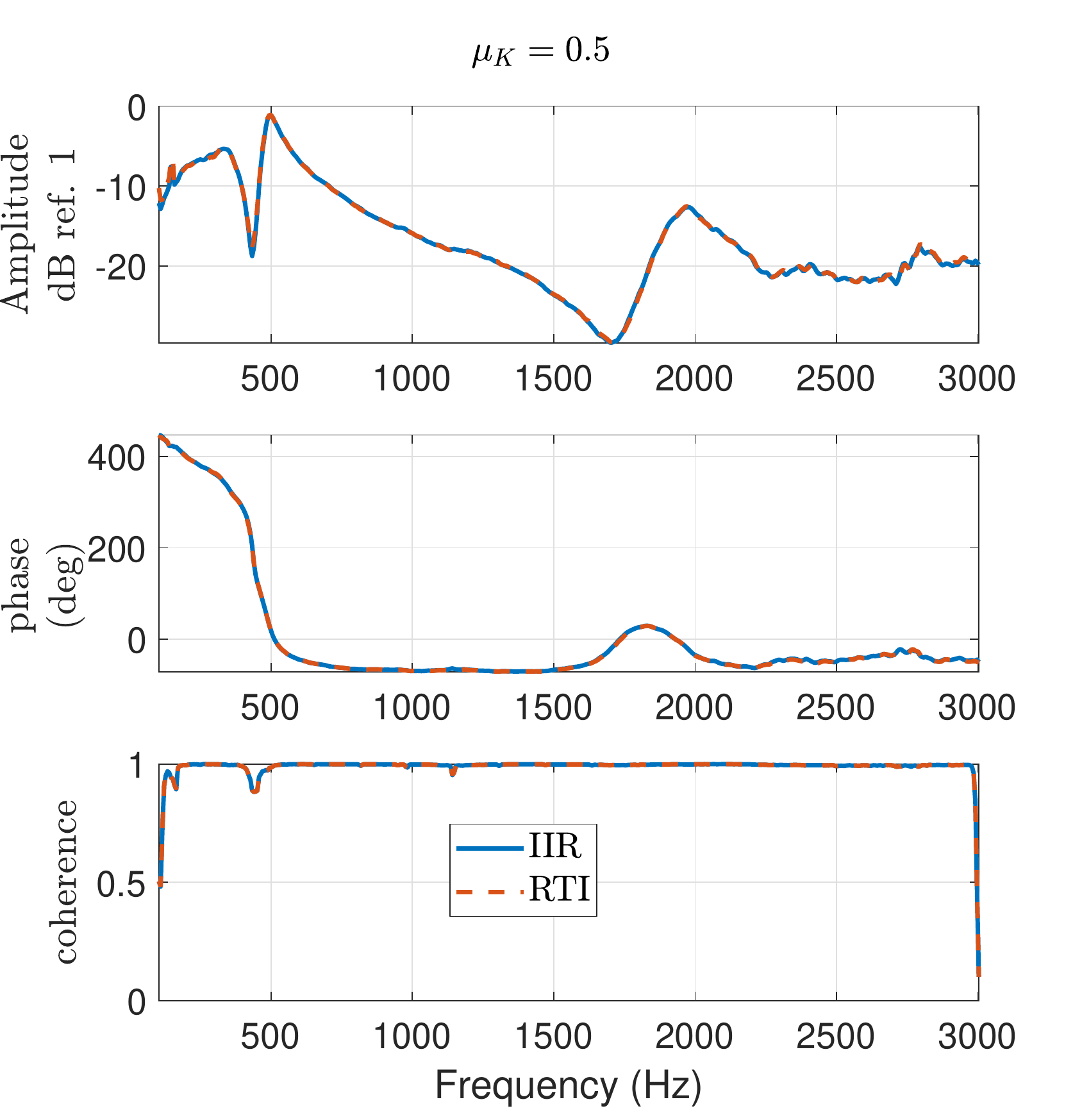}
			\centering
			\caption{}
			\label{fig:mobility_IIRvsRTI_0.5muK}
		\end{subfigure}
		\hspace{0.5 cm}
		\begin{subfigure}[ht!]{0.45\textwidth}
			\centering
			\includegraphics[width=\textwidth]{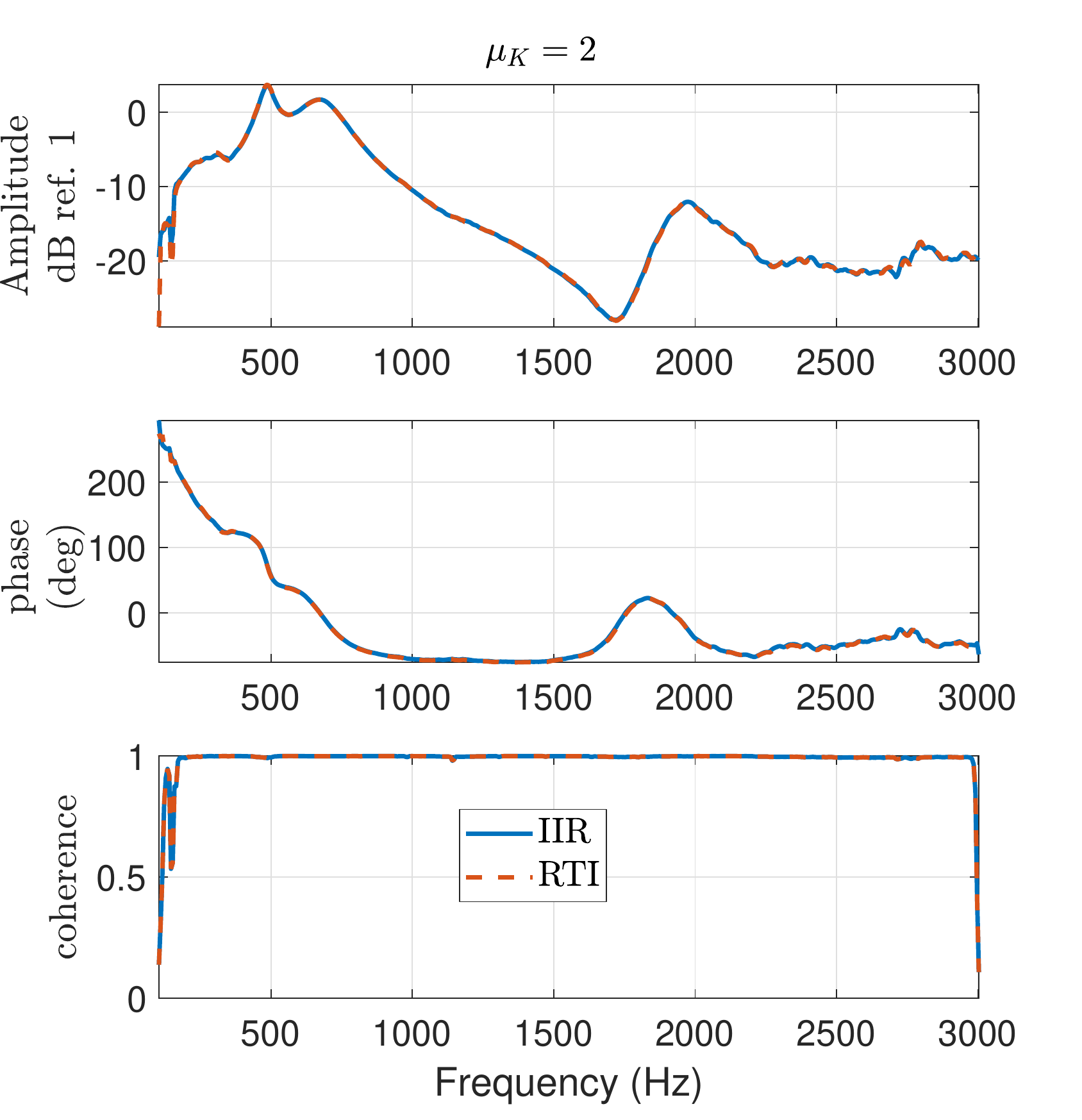}
			\caption{}
			\label{fig:mobility_IIRvsRTI_2muK}
		\end{subfigure}
		\caption{Mobility obtained by targeting a linear SDOF dynamics of the EA with $\mu_K=0.5$ \textbf{(a)} and $\mu_K=2$ \textbf{(a)}, $\mu_M=1$ and $R_{at}=\rho_0c_0$, by either the IIR (solid blue line) or the RTI (dashed red line).}
		\label{fig:mobility_IIRvsRTI_Varying_muK}
	\end{figure}

\begin{figure}[ht!]
		\centering
		\begin{subfigure}[ht!]{0.45\textwidth}
			\centering
			\includegraphics[width=\textwidth]{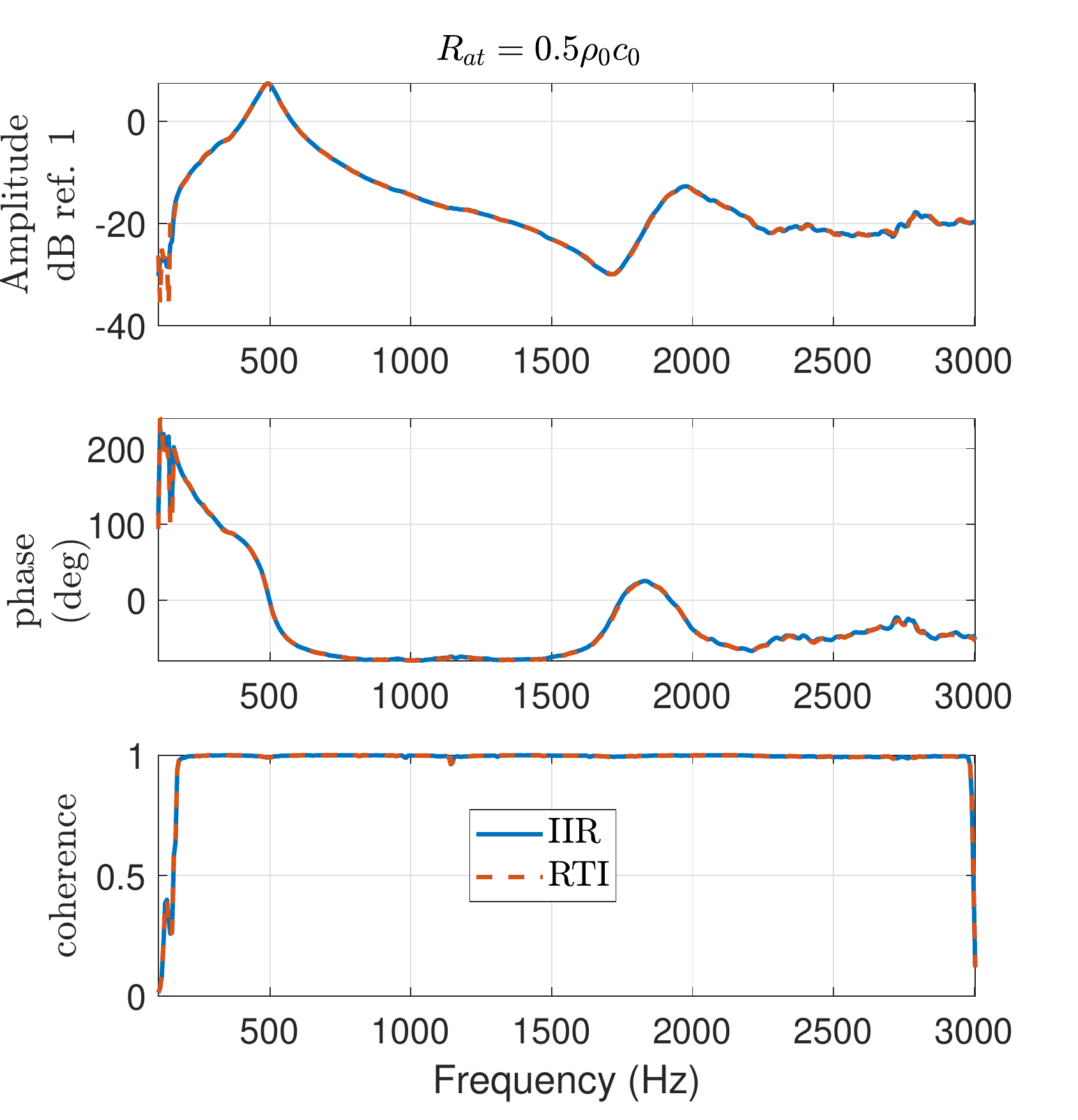}
			\centering
			\caption{}
			\label{fig:mobility_IIRvsRTI_0.5RatRho0c0}
		\end{subfigure}
		\hspace{0.5 cm}
		\begin{subfigure}[ht!]{0.45\textwidth}
			\centering
			\includegraphics[width=\textwidth]{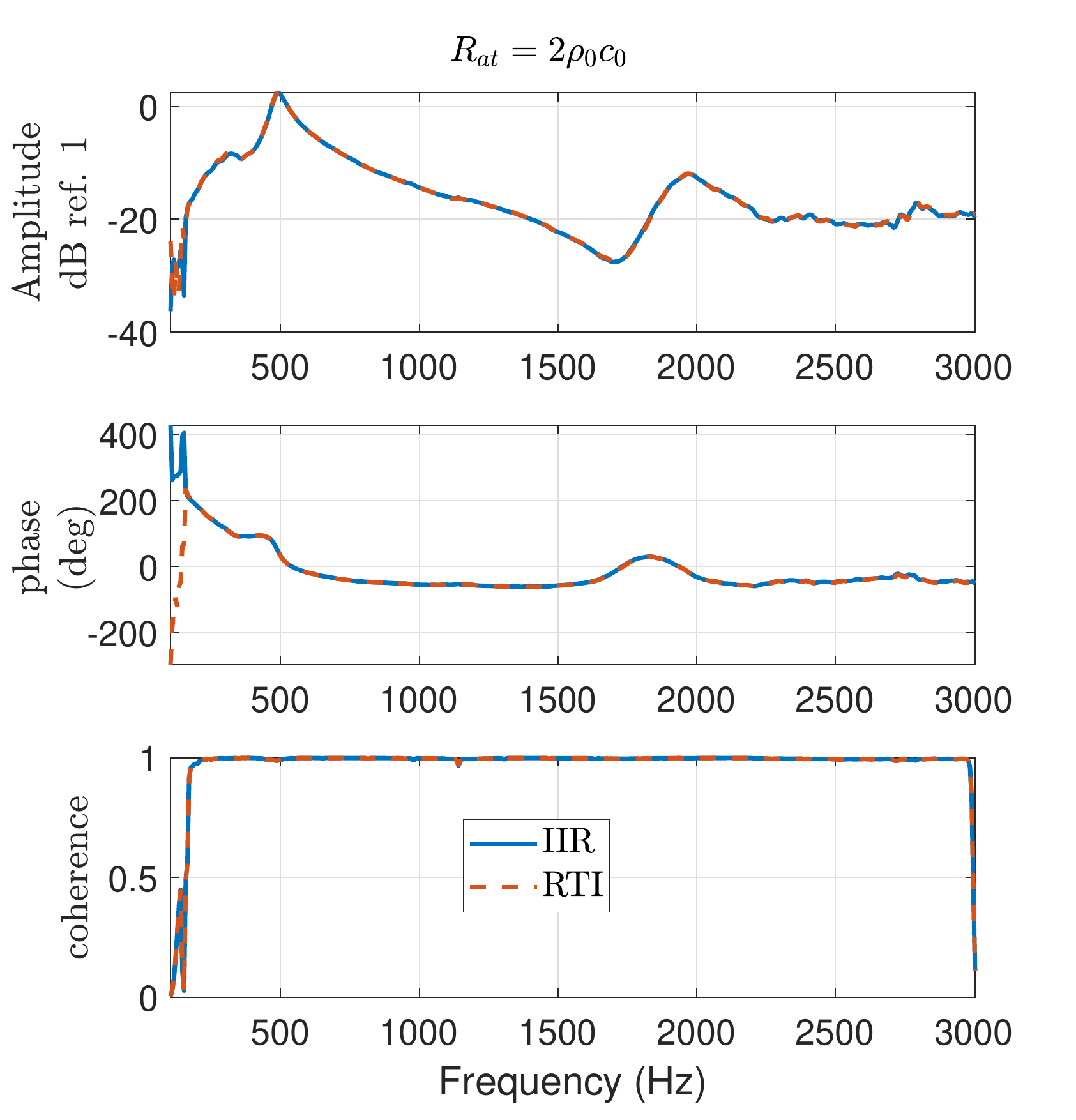}
			\caption{}
			\label{fig:mobility_IIRvsRTI_2RatRho0c0}
		\end{subfigure}
		\caption{Mobility obtained by targeting a linear SDOF dynamics of the EA with $R_{at}=0.5\rho_0c_0$ \textbf{(a)} and $R_{at}=2\rho_0c_0$ \textbf{(a)}, and $\mu_M=\mu_K=1$, by either the IIR (solid blue line) or the RTI (dashed red line).}
		\label{fig:mobility_IIRvsRTI_Varying_RatRho0c0}
	\end{figure}

	Below we report the mobilities of the EA controlled by either the IIR or the RTI control strategy with linear target dynamics. In Figure \ref{fig:mobility_IIRvsRTI_Varying_muM} both the target mass and stiffness are equally varied to half (Figure \ref{fig:mobility_IIRvsRTI_0.5muMmuK}) and twice (Figure \ref{fig:mobility_IIRvsRTI_2muMmuK}) the open circuit values. Figure \ref{fig:mobility_IIRvsRTI_Varying_muM} and \ref{fig:mobility_IIRvsRTI_Varying_muK} shows the single variation of the target mass and stiffness respectively, while Figure \ref{fig:mobility_IIRvsRTI_Varying_RatRho0c0} shows the variation of the target resistance term. These results demonstrate the total equivalence of the two strategies IIR and RTI to achieve tunable linear target dynamics.
	
\end{appendices}

\FloatBarrier

\bibliographystyle{apsrev}
\bibliography{biblio_manuscript_2021_2}

\end{document}